\documentclass[a4paper,fleqn]{cas-dc}

\usepackage[numbers, sort&compress]{natbib}

\usepackage[utf8]{inputenc}

\def\tsc#1{\csdef{#1}{\textsc{\lowercase{#1}}\xspace}}
\tsc{STM}
\tsc{STS}
\tsc{PLD}
\tsc{vdW}
\begin{document}
\let\WriteBookmarks\relax
\def\floatpagepagefraction{1}
\def\textpagefraction{.001}

% Short title
\shorttitle{STM study of Bi$_2$Te$_3$ thin films grown by PLD}

% Short author
\shortauthors{NI Fedotov {\it et~al.}}

% Main title of the paper
%\preprint{ver 0}
%STM study of PLD deposition of thin Bi2Te3films from tellurium-reach target
\title[mode = title]{Scanning tunneling microscopy of Bi$_2$Te$_3$ films prepared by pulsed laser deposition: from nanocrystalline structures to van der Waals epitaxy}
%{Van der Waals Epitaxy of Сrystalline Films of Bi$_2$Te$_3$ by PLD: STM Study}
%\thanks{A footnote to the article title}%
 \author[1]{N.I. Fedotov}[orcid=0000-0001-6145-109X]
 
\author[1]{A.A. Maizlakh}
% \affiliation%[Also at ]%{Physics Department, XYZ University.}%Lines break automatically or can be forced with \\
 
 \author[1]{V.V. Pavlovskiy}
 
 \author[2]{G.V. Rybalchenko}
 
 \author[1,3]{S.V. Zaitsev-Zotov}[ orcid=0000-0002-5942-1323]
 \ead{SerZZ@cplire.ru}
\cormark[5]
 % \email{Second.Author@institution.edu}
 
% \affiliation{Kotelnikov Institute of Radioengineering and Electronics of Russian Academy of Sciences, Mokhovaya 11, bld. 7, Moscow, 125009 Russia }

\affiliation[1]{organization={Kotelnikov Institute of Radioengineering and Electronics of Russian Academy of Sciences},
    addressline={Mokhovaya 11, bld. 7}, 
    city={Moscow},
    % citysep={}, % Uncomment if no comma needed between city and postcode
    postcode={125009}, 
    % state={},
    country={Russian Federation}}
    
\affiliation[2]{organization={P. N. Lebedev Physical Institute},
    addressline={Leninsky pr., 53}, 
    city={Moscow},
    % citysep={}, % Uncomment if no comma needed between city and postcode
    postcode={119991}, 
    % state={},
    country={Russian Federation}}
    
\affiliation[3]{organization={National Research University Higher School of Economics},
    addressline={20 Myasnitskaya Street}, 
    city={Moscow},
    % citysep={}, % Uncomment if no comma needed between city and postcode
    postcode={101000}, 
    % state={},
    country={Russian Federation}}

\date{\today}% It is always \today, today,
             %  but any date may be explicitly specified

\begin{abstract}
Bi$_2$Te$_3$ is a material with high efficiency of thermoelectric energy conversion. Recently, it was also recognized as a topological insulator, and is often used as the basis for creation of other types of topological matter. Pulsed laser deposition (PLD) is widely considered as a simple method for growing of multicomponent films, but not as a tool for van der Waals epitaxy. We demonstrate here that the van der Waals epitaxy of Bi$_2$Te$_3$ is indeed impossible in vacuum PLD, but is possible in the presence of a background gas, which is confirmed by the results of scanning tunneling microscopy and spectroscopy studies. Results of {\it ab initio} calculations reproduce tunneling spectra of the first three terraces of epitaxial films of Bi$_2$Te$_3$. In addition, an unusual hexagonal superstructure resembling a charge-density wave is observed in overheated films.
\end{abstract}

%\begin{highlights}
%\item Pulsed laser deposition of Bi$_2$Te$_3$ in vacuum damages the substrate preventing thereby van der Waals epitaxy

%\item Van der Waals epitaxy of Bi$_2$Te$_3$ by pulsed laser deposition is possible in argon atmosphere

%%\item The bottom layer of epitaxial  Bi$_2$Te$_3$ films contains an additional Bi bilayer 

%\item Tunneling spectra of the first three terraces of the film correspond to calculated local density of states of 1, 2 and 3 quintuple layers respectively

%\item Annealing of Bi$_2$Te$_3$ films at temperatures $\gtrapprox 250$ \textdegree C in ultra high vacuum conditions results in development of new hexagonal superstructure with 4 nm period
%\end{highlights}

\begin{keywords}
\sep Bi$_2$Te$_3$
\sep pulsed laser deposition 
\sep scanning tunneling microscopy 
\sep scanning tunneling spectroscopy 
\sep thin films 
\sep van der Waals epitaxy
\end{keywords}

%\keywords{Suggested keywords}%Use showkeys class option if keyword
                              %display desired
\maketitle

%\tableofcontents

%\input{intro_v5}
\section{Introduction}

Bi$_2$Te$_3$ is a layered material consisting of  covalently bonded Te-Bi-Te-Bi-Te  quintuple layers (QL) stacked along the c-axis to form a rhombohedral crystal. The quintuple layers are interconnected by a relatively weak van der Waals (vdW) interaction. This compound is widely known as a material with high efficiency of thermoelectric energy conversion \cite{satterthwaite1957electrical}. In the last decade it has been also intensively studied as a topological insulator \cite{chen2009experimental} -- a novel type of solid state material and part of a larger class of topological materials, 
which includes magnetic topological insulators, various types of Weyl and Dirac semimetals, and topological superconductors.
In general, their physical properties 
are determined by the presence of certain symmetries describing their band structures \cite{Yan_2012}. Topological insulators are characterized by the presence of a bulk energy gap and topologically protected edge or surface states.
Presently, topological insulators Bi$_2$Te$_3$ and Bi$_2$Se$_3$ remain the most popular topological materials and even are often used as  the basis  for creation of other types of topological matter. For instance, by  proper doping they can be transformed  into topological superconductors \cite{TI_doping} or a magnetic topological insulator \cite{Chang167}.

The structure of films for thermoelectric applications and of those suitable for the use of their topological properties must satisfy different requirements. The films of interest for thermoelectric applications  should contain a large number of grain boundaries that prevent phonon propagation. Meanwhile, for topological applications, high quality continuous films or isolated islands are the most appealing.

In thin films of topological materials, a number of exotic states can be realized, such as a quantum spin Hall state, an axion insulator, an anomalous quantum Hall state, as well as numerous effects at the boundaries of films, islands, and grains \cite{Culcer_2020}.  
One way to obtain thin films of layered topological materials would be mechanical cleaving along weak vdW bonds using micromechanical cleavage \cite{Novoselov10451}. This technique is usually used for obtaining clean surface of bulk crystals, and also was heavily utilized in early graphene studies. However, it is poorly scalable for industrial usage. 
Therefore more and more attention has been paid to methods for growing thin films of vdW materials, such as vdW epitaxy \cite{vdW_epitaxy,vdW_epitaxy_on_graphene}. Due to very low surface energy of vdW materials, it allows thin films to be grown without meeting the requirements for lattice matching between substrate and film, especially if substrate is also a vdW material.

One way to achieve vdW epitaxy is to employ widely used molecular beam epitaxy (MBE) and use a substrate without dangling bonds. To date, there exist a number of reports detailing various ways to grow Bi$_2$Te$_3$ (see for instance
\cite{KRUMRAIN2011115,wang2011topological,Roy_2013,Liu_2011,Harrison_2013,Macedo_2015,Rodrigues_2020}). In all mentioned works the substrate was heated to a temperature $T_s$=170--350 \textdegree C during the deposition process. The heating provides surface adatom mobility and allows the formation of large single-crystal films. However, it also results in tellurium deficiency due to  high vapor pressure of Te at those temperatures. An excess tellurium flux is required to maintain stoichiometry of Bi$_2$Te$_3$. The typical beam equivalent pressure (BEP) ratio of Te$_2$ flux to Bi flux was around 20 in most cases. It was also shown that a higher $T_s$ requires higher BEP ratio.

One notable example of MBE growth of Bi$_2$Te$_3$ film  is described in Ref.~\cite{Rodrigues_2020}.
In this work, one effusion cell was loaded with Bi$_2$Te$_3$ instead of pure Bi. The resulting BEP for Te$_2$ flux to Bi$_2$Te$_3$ flux is given as 2, which corresponds to Te$_2$ flux to Bi flux ratio of 1.75.  The growth was performed on highly oriented pyrolytic graphite (HOPG). Such a combination of low BEP ratio with the high surface adatoms mobility substrate provides the lowest optimal $T_s = 170$ \textdegree C. In addition, obtained films were as thin as 1.4 QL (given that 1 QL is roughly 1 nm, the thickness therefore is 1.4 nm). The results could be considered  a fine example of vdW epitaxy by MBE.

Aside from MBE, another promising method to achieve vdW epitaxy is pulsed laser deposition (PLD) \cite{PLD_review}. It is a relatively easy way to produce multi-component films. However, early works \cite{Dauscher_1996,Bailini2007} regarding PLD of Bi$_2$Te$_3$ show that deposition of films in vacuum $< 10^{-7}$ Torr results in Te deficiency. The issue persists even when a substrate is at room temperature. The effect is largely attributed to Te plume expansion and plume-laser beam interaction. A consistent way to prevent Te loss is to introduce inert gas into the chamber during deposition, as shown by Bailini, {\it et al.} \cite{Bailini2007}. This and later works use  Ar at a pressure of $1.5 \times 10^{-2}$ - 3 Torr
\cite{Bailini2007,Obara_2009,Wudil_2020,Zhang_2012,Yu_2013,Le_2014,Liao_2019,Jian_2017}
or He at $2 \times 10^{-5}$--$2 \times 10^{-3}$ Torr
\cite{Tasi_2015} for this purpose.
Optimal substrate temperature varies between 200 to 300 \textdegree C. In \cite{Bailini2007} it is also shown that correct stoichiometry of the film can be obtained due to thermal evaporation of extra Te if a target with Te excess is used.

This work presents the results of STM studies of films grown in a two-step process both in vacuum and argon atmosphere at different substrate temperatures and further annealing in vacuum. As a substrate in our experiments, we used highly oriented pyrolytic graphite (HOPG), which has the conductivity required for using STM, the same rotational symmetry, no dangling bonds and therefore being vdW material appropriate for vdW epitaxy.
The results obtained indicate the possibility of controlling the film morphology by choosing deposition and subsequent annealing conditions and provide recipes for fabrication of films with the surface morphology required.  We  demonstrate also that PLD in vacuum results in substrate damage  and therefore cannot be used for vdW epitaxy. Instead, PLD at a moderate inert gas pressure followed by annealing at temperatures 150-200\textdegree C provides vdW epitaxy of ultrathin crystalline films of Bi$_2$Te$_3$.

\section{Methods}

A home-made PLD setup connected to the Omicron LT SPM  by a vacuum transport channel through an additional technological chamber (XP chamber) was used for film deposition.
Laser radiation was focused into a spot 1 mm in diameter on a target with a typical size of 10 × 15 mm. The target was scanned using a computer-controlled lens positioning system providing typical  scanning rate of 100 mm/min, which ensured that there was no overlap between the light spots. The second harmonic ($\lambda =532$~nm) of a neodymium laser working at a repetition rate of 0.2 - 0.5 Hz with 7 ns pulse duration and  80 - 82 mJ pulse energy was used.  The target-substrate distance was 5 cm. The nominal deposition rate was 2.5 \AA /min. The amount of deposited material was controlled by a quartz microbalance located next to the substrate holder. The base vacuum in the deposition chamber was $1-2 \times 10^{- 8} $ Torr and it deteriorated to $\sim 10^{- 7}$~Torr in the process of deposition.
Prior to every deposition, HOPG substrate was cleaved {\it ex situ}  and annealed in vacuum at $T = 430$ \textdegree C during at least 1 hour.
The temperature of the substrate in the PLD chamber was measured with a thermocouple calibrated with a thermistor glued by conductive carbon paint at substrate position, the calibration curves for vacuum and argon being different. Post-deposition annealing was carried out in ultra-high vacuum conditions in the Omicron XP chamber. The temperature regime during deposition and subsequent annealing was maintained automatically. STM/STS study was performed in LT STM chamber at pressure $2\times 10^{-11}$~Torr and temperature 78~K. Typical set points for STM images and STS spectra collection are $\pm 0.5$~V and 0.2 nA.

The elemental composition of the films was investigated using SEM-EDS analysis performed on films with a nominal thickness of 100-150 nm in order to reduce the contribution of the substrate. JSM-7001F scanning electron microscope (JEOL) with an energy-dispersive X-ray analyzer (INCAx-act Oxford Instruments) and INCA software was used. The EDX spectra were obtained using 20~kV accelerating voltage and beam current of about $~$0.07~mA. The average sample composition was calculated after at least ten point measurements during 1800 sec each. It is important to note that the distribution of Bi and Te in each thin film was additionally analyzed by element mapping during several hours. 
 
The electronic band structure was calculated using density functional theory (DFT) with the local-density
approximation (LDA) and spin-orbit interaction taken into account. A preliminary structure optimization was performed for each structure using 8x8x1 $\vec{K}$-mesh.

For the STM spectra calculations  Tersoff-Hamann approximation \cite{TersoffHamann1985} was used, according to which STM differential tunneling conductance is proportional to the local density of
states (LDOS) $N(\vec{R},E)$
\begin{equation}
    N(\vec{R},E)=\sum _{\nu }\left|\psi _{\nu } \left(\vec{R}\right)\right| \delta \left(E-E_{\nu } \right),
    \label{eq:Es}
\end{equation}
where $\psi _{\nu }(\vec{R})$ is the wave function at the point  of the tip location
 $\vec{R}$ and $E_{\nu } $ is the corresponding eigenenergy.

 Tip location was chosen at a proper distance from the surface of the slab in the  vacuum gap, where wave functions demonstrate an exponential decay. Pseudo wave functions
   in this area coincide with true electron wave functions and the influence of slab's opposite surface is rather low.
    Dirac's $\delta $-function was approximated by Lorentz's function with 40 meV width, which corresponds to the voltage
    resolution of STM. Pseudo wave functions were calculated using optimized slab crystal structures on 24x24x1 $\vec{K}$-mesh.

\section{Preliminary notes}

 The choice of the target composition 
 %in the PLD method greatly 
  may affect  the deposition results. When using the MBE method, sources providing 8-20 time larger Te over Bi fluxes are required at substrate temperature 200-600 \textdegree C \cite{KRUMRAIN2011115,chen2011MBE}. In the PLD method realized in vacuum conditions some tellurium  loss also occurs, so targets with extra tellurium content over  stoichiometry are also required \cite{Bailini2007}.
  From the Bi-Te phase diagram \cite{phasediag} we expect the films at more than 60 at. percent Te to consist of two phases: Te and Bi$_2$Te$_3$. Given the high vapor pressure of Te we expect it to evaporate upon annealing, leaving only Bi$_2$Te$_3$ behind.

\begin{figure}[h]
\center{\includegraphics[width=7cm]{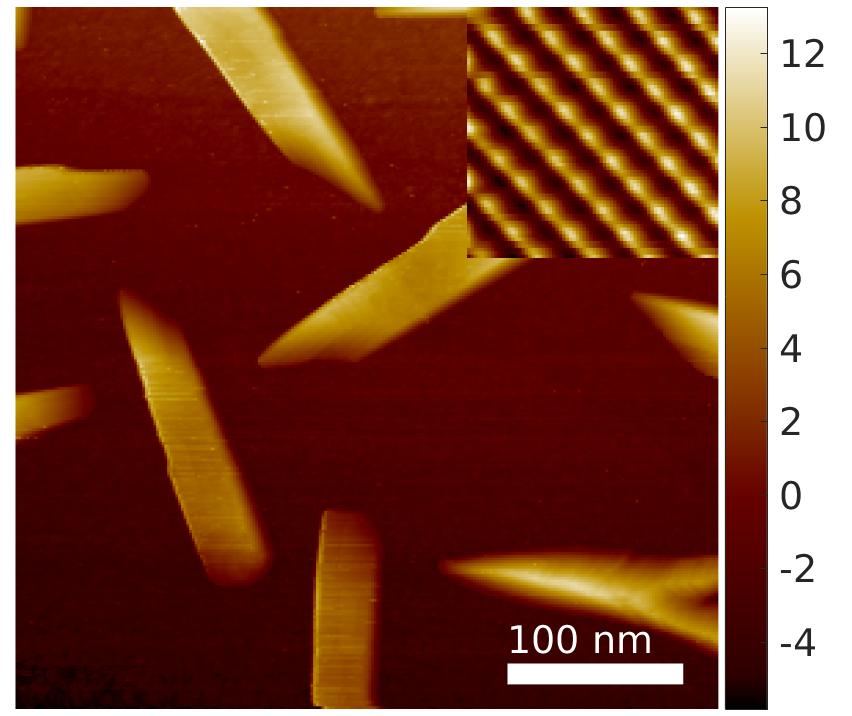}}

\caption{\label{fig:Te100}
STM image of a Te film deposited at substrate temperature 120 \textdegree C in vacuum. Inset shows 
a fragment of atomically resolved surface of the central crystallite.}
\end{figure}

We have undertaken a test deposition of tellurium to study the effect of annealing. The results are shown in Fig. \ref{fig:Te100}. Tellurium deposited at a substrate temperature of 120~\textdegree C aggregates in large crystals with the atomic structure corresponding to $\beta$-Te \cite{Te_STM} (see the inset) whereas it forms a continuous amorphous film if it is deposited at 60~\textdegree C. Amorphous film becomes polycrystalline upon annealing at 120-175~\textdegree C and evaporates from the HOPG substrate as a result of annealing at $T\gtrsim 240$\textdegree C.  Thus the amount of  tellurium can be adjusted to proper stoichiometry of the film by choosing the substrate temperatures during deposition and subsequent annealing below 300~\textdegree C. In the present work, we used a target with an atomic ratio of components Bi/Te = 1/5 which allows us to deposit films with both over and under stoichiometry amount of tellurium.

The phase of our interest Bi$_2$Te$_3$ is the most stable one as it has the minimum formation energy among other phases. The other ones, described by the formula (Bi$_2)_n$(Bi$_2$Te$_3)_m$, consist of alternation of Bi$_2$Te$_3$ quintuple layers with bismuth bilayers. Some examples are  BiTe \cite{Yamana1979} (which is also a topological insulator \cite{Eschbach_2017}), Bi$_4$Te$_3$ \cite{Yamana1979} {\it etc.} They are formed when the  content of tellurium in the system is less than 60 percent. 

Phase identification of ultra-thin Bi-Te films by STM can be done by study of atomic structure of film surface, heights of terrace steps and scanning tunneling spectroscopy (STS) spectra. The main phases that may appear are bilayers of Bi, quintuple layers of Bi$_2$Te$_3$, septuple layers (SL) of Bi$_4$Te$_3$, with thicknesses 0.34, 1.01 and 1.35~nm respectively. All these phases have hexagonal symmetry of surface atomic structure with very close interatomic distance values which are hardly distinguishable by STM, but their energy structures are different.

\section{Results}
\subsection{{\it ab initio} calculations}

STM spectra were calculated from the results of {\it ab-initio} band structure calculations.
In order to build crystal structures which are suitable for the surface electronic structure investigation
we used unit cells of Bi${}_{2}$Te${}_{3}$ and Bi${}_{4}$Te${}_{3}$ compounds with hexagonal axes as basic
structures. Structure parameters were taken from \cite{MatProj}. Slab structures with the boundaries required were 
cut from the base structures along the $z$ plane with vacuum gaps added.

The following slab structures were used:

(i) one, two and three Bi${}_{2}$Te${}_{3}$ QLs with Te layer surface and  the vacuum gap above;

(ii) three Bi${}_{2}$Te${}_{3}$ QLs cleaved between 2-nd and 3-rd layer of the central QL and the vacuum gap in the break;

(iii) one SL of Bi${}_{4}$Te${}_{3}$ with Bi bi-layer at the surface and the vacuum gap above;

\begin{figure*}
\includegraphics[width=3.7cm]{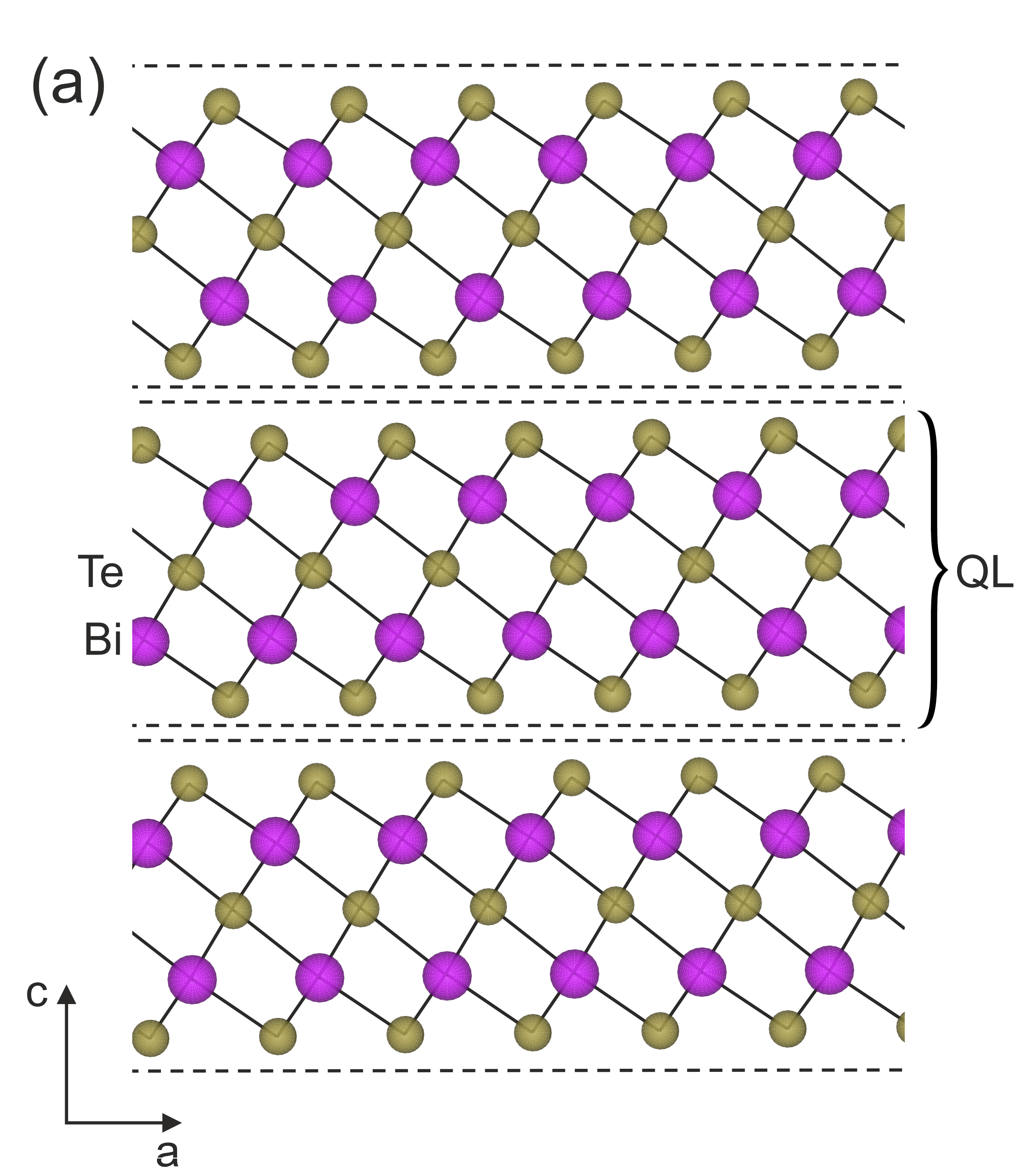}
\includegraphics[width=3.5cm]{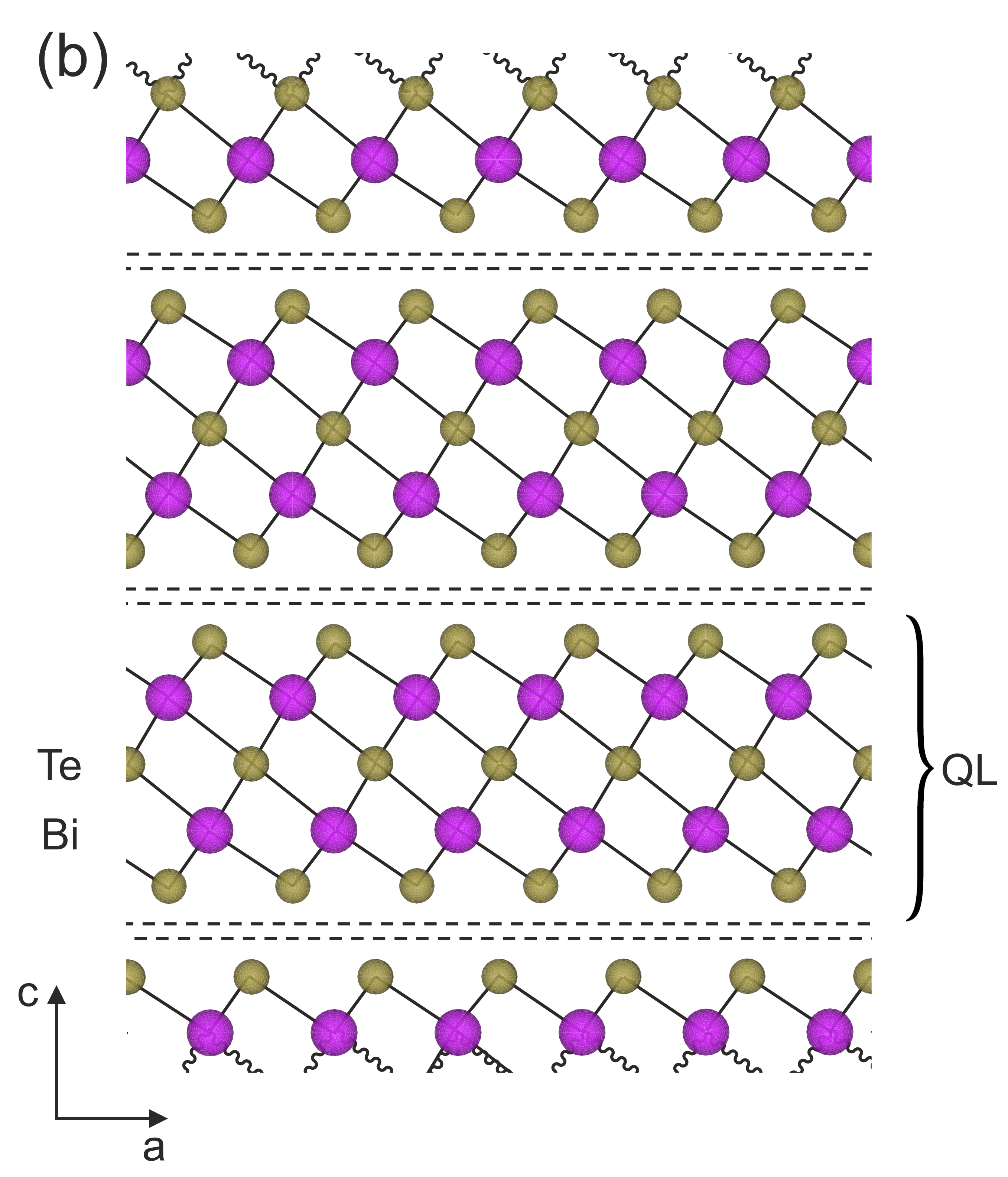}
\includegraphics[width=3.5cm]{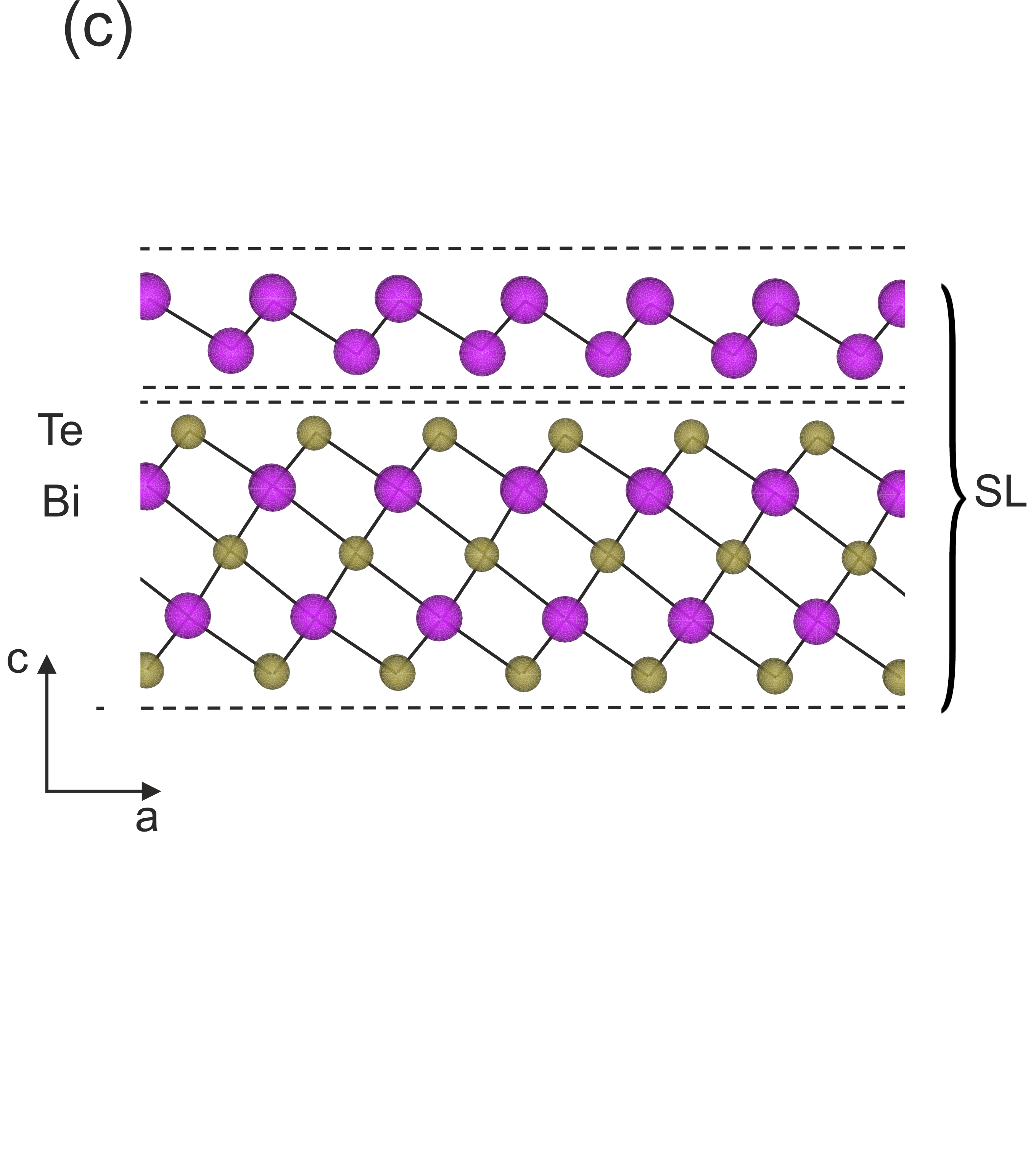}
\includegraphics[width=5.7cm]{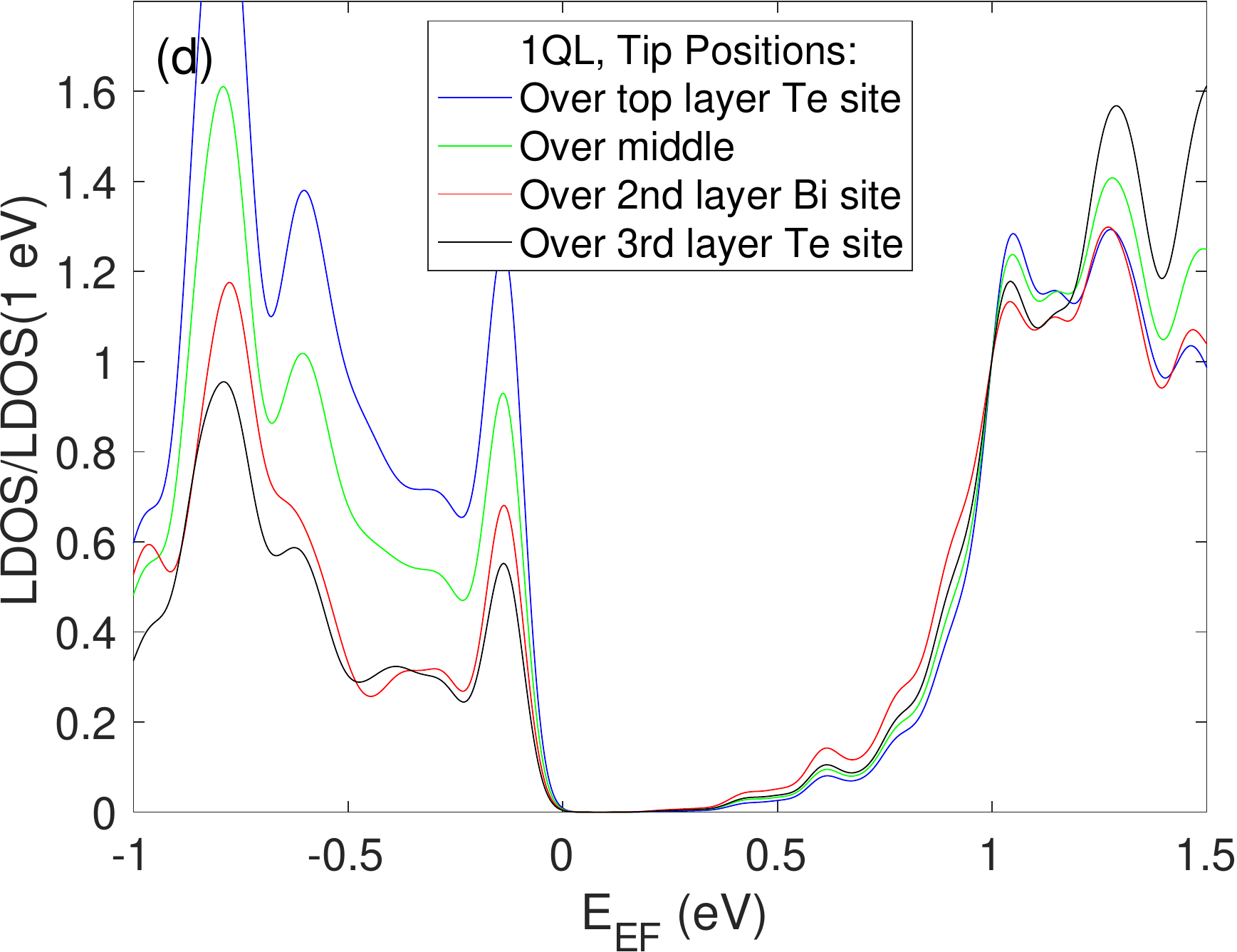}

\includegraphics[width=5.7cm]{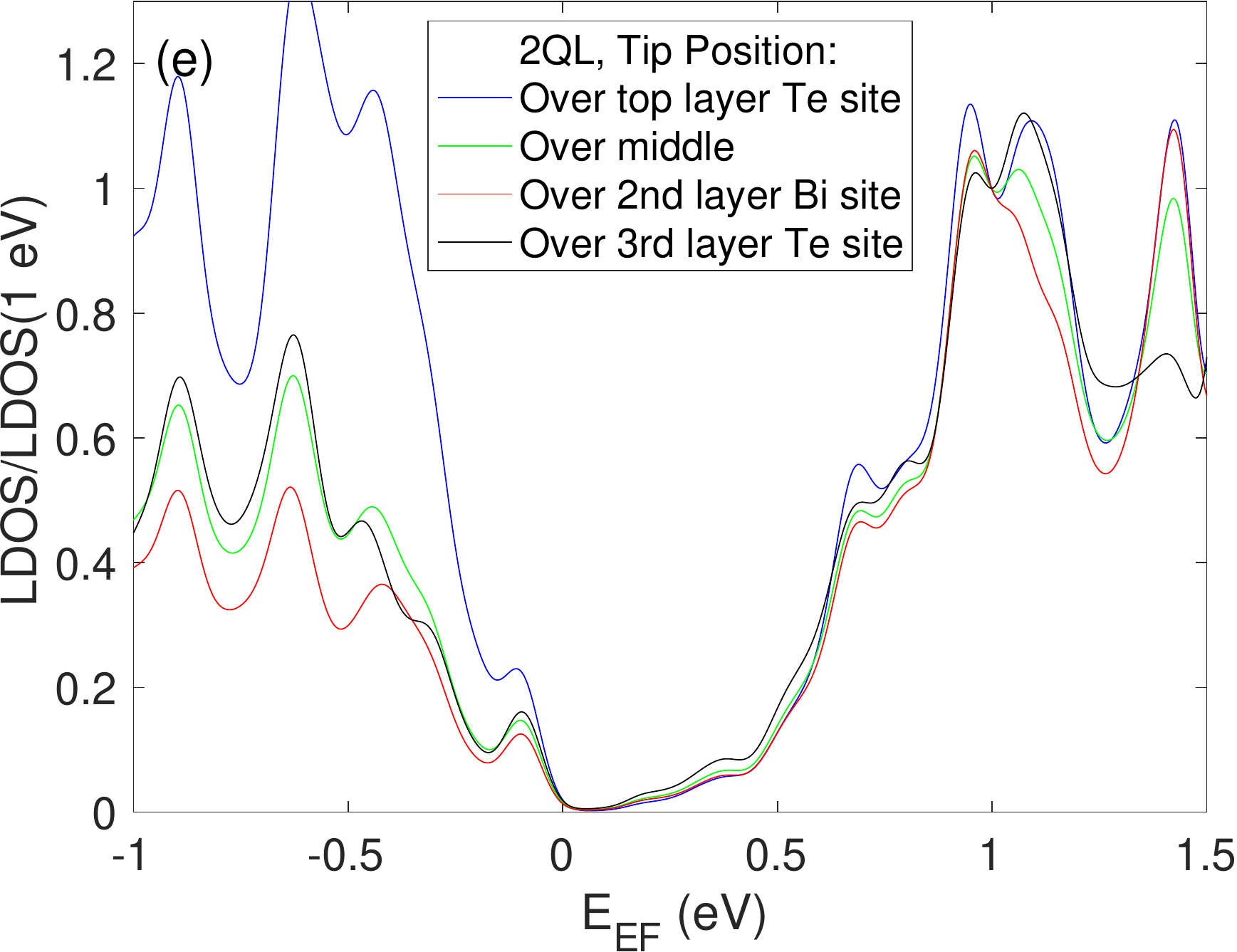}
\includegraphics[width=5.7cm]{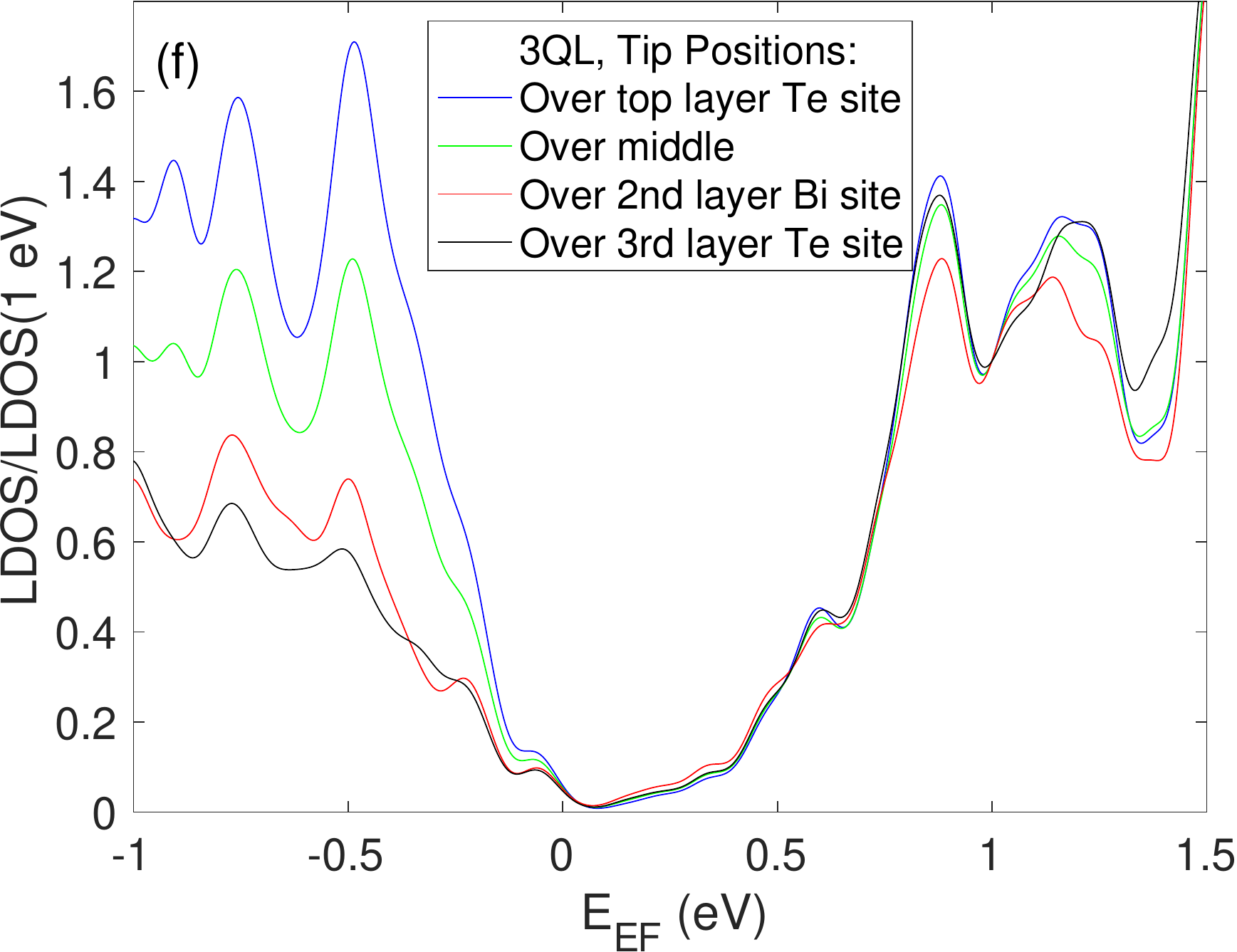}
\includegraphics[width=5.7cm]{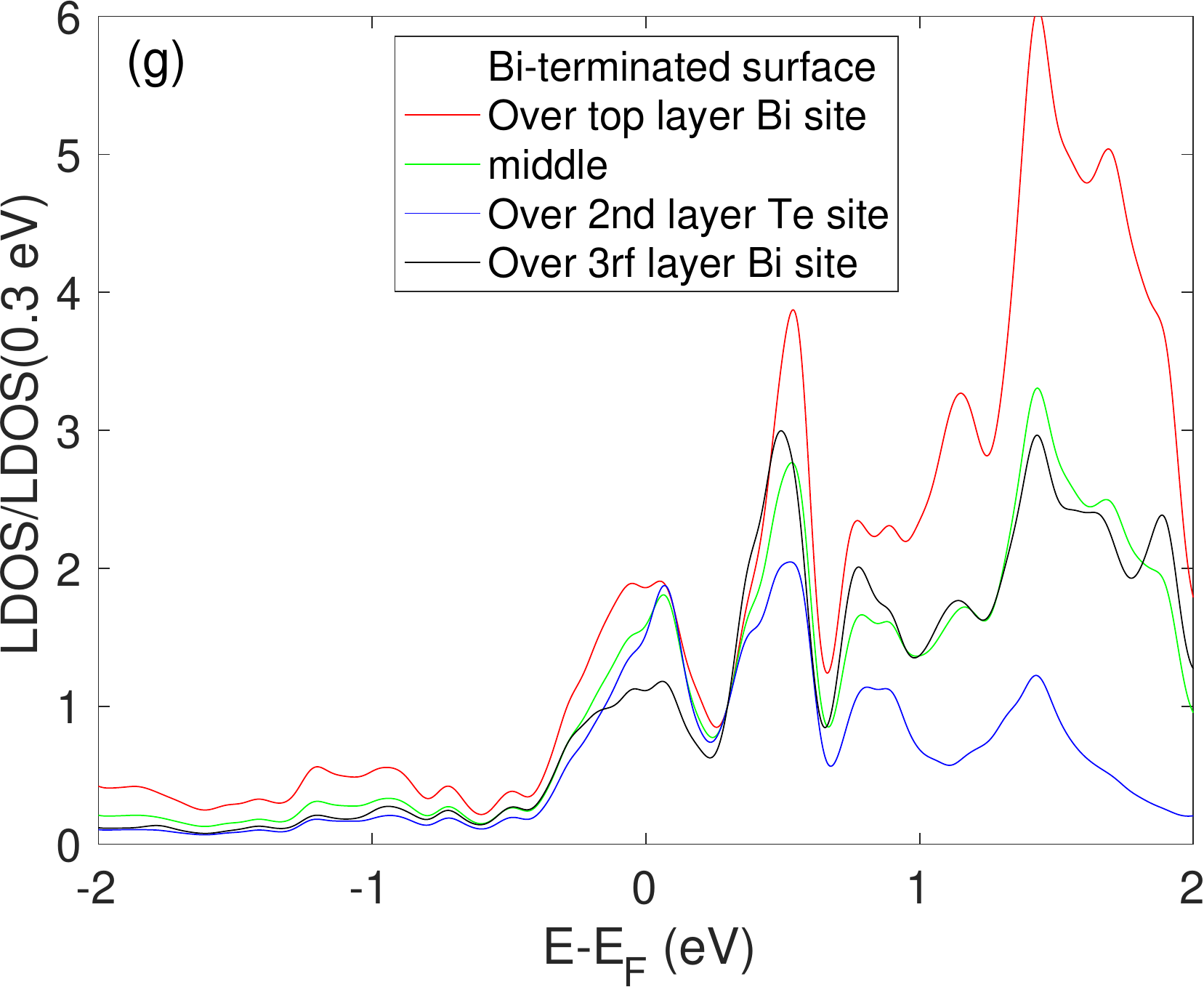}

\includegraphics[width=5.7cm]{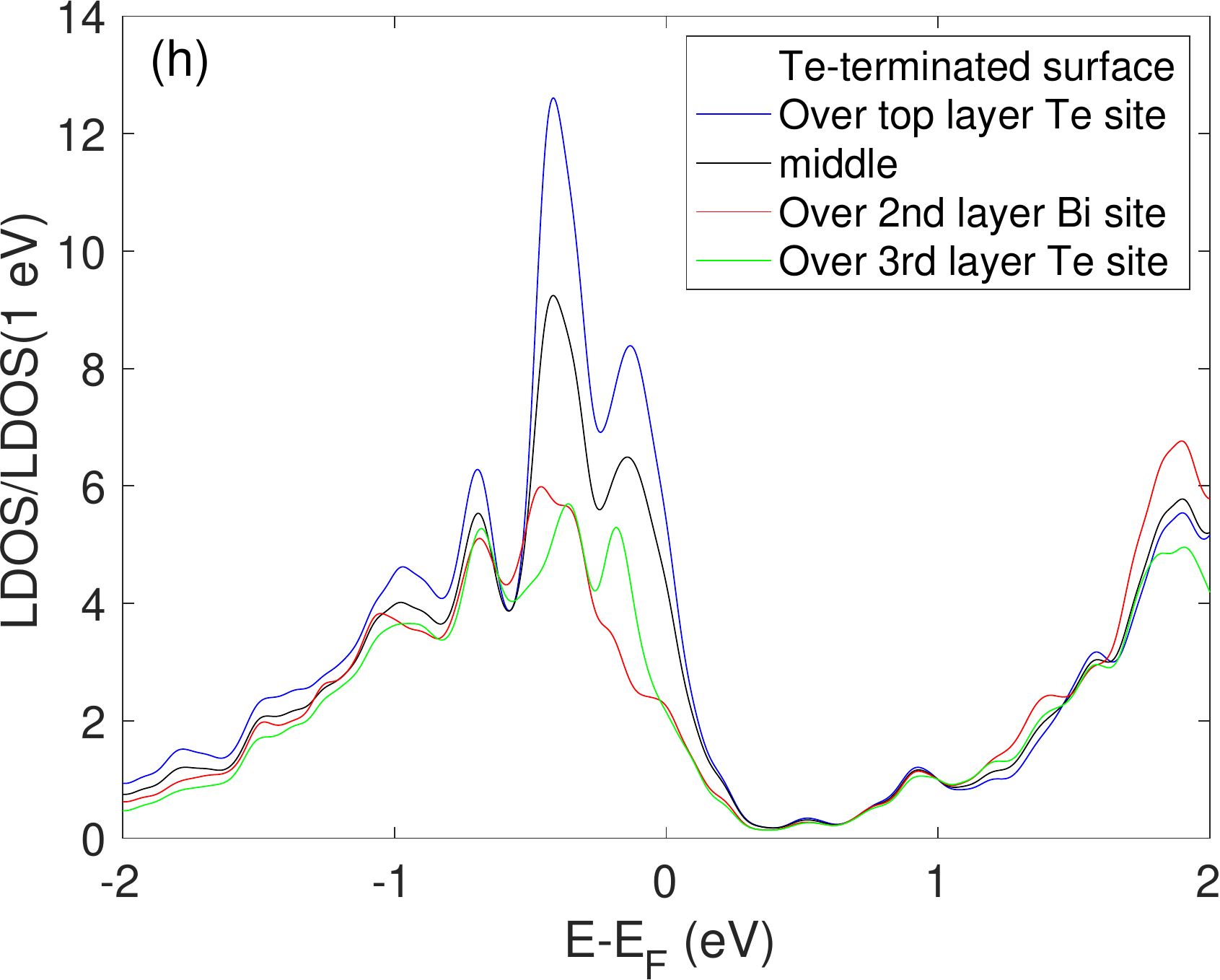}
\includegraphics[width=5.7cm]{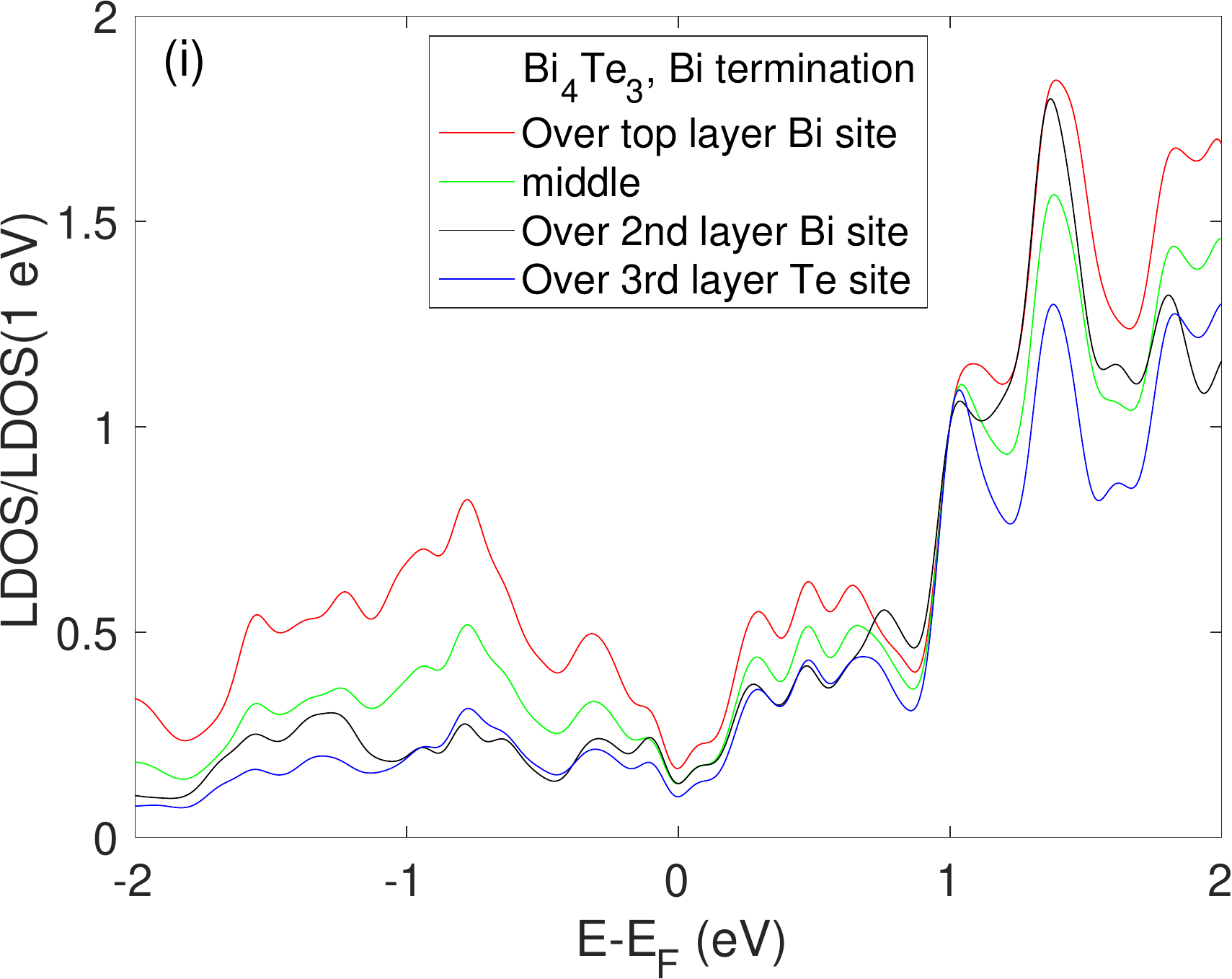}
\includegraphics[width=5.7cm]{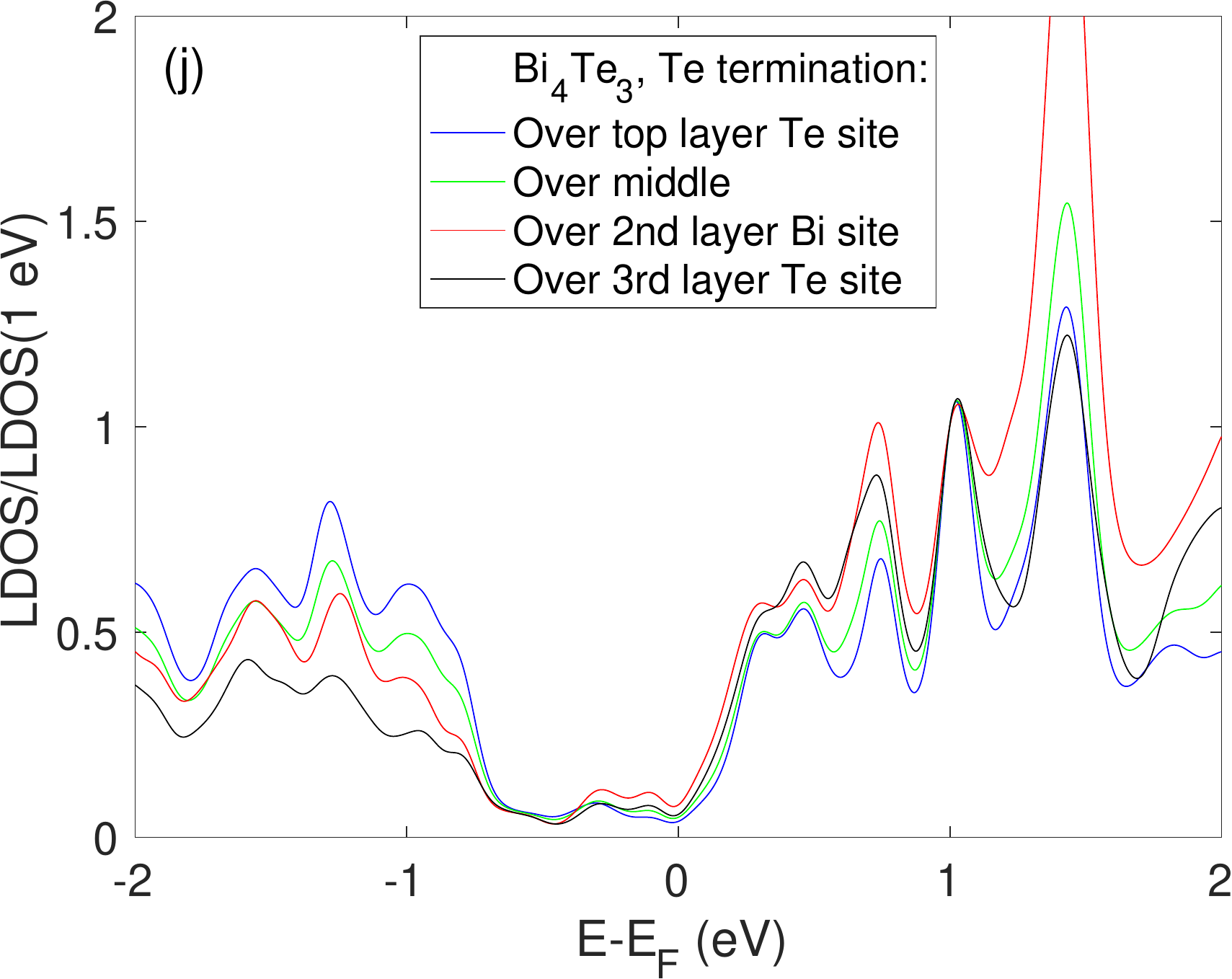}

\caption{\label{fig:STS} (a-c) Slab structures used for  calculations: (a) 3QL slab; (b) a slab with abnormal Bi-terminated and Te-terminated surfaces; (c) 1SL of Bi$_4$Te$_3$ slab. Chemical bonds are indicated by solid lines for covalent, double dashed lines for vdW, single dashed lines for broken vdW and wavy lines for broken covalent bonds. The tip Z-position is 3 \AA\ over the upper surface.
 LDOS of: (d) 1QL Bi$_2$Te$_3$; (e) 2QL  Bi$_2$Te$_3$; (f) 3QL  Bi$_2$Te$_3$. (g,h) LDOS of Bi$_2$Te$_3$ slab with abnormal terminated surfaces: (g) Bi-terminated surface; (h) Te-terminated surface. LDOS of 1SL of Bi$_4$Te$_3$ for (i) Bi- and (j) Te-terminated surfaces.}
\end{figure*}

Structures (ii) and (iii) correspond to different types of Bi$_2$Te$_3$ surfaces with abnormal terminations 
 which can also be observed experimentally \cite{termination}.
The geometry of the three QL unit cell with a break in the middle was chosen for the structure (ii)
to exclude electrical polarization of the unit cell. It is shown as a continuous slab with abnormal terminated surfaces in fig. 2b below. Vacuum gap width of 16 \AA\ and 3 \AA\  distance of the tip and the surface were used for all structures. Main expected structures used for calculations are presented at Fig. ~\ref{fig:STS}(a-c). These pictures were prepared with the usage of Vesta program \cite{MommaIzumi2011}.
    
Fig.~\ref{fig:STS}(d-j) shows local density of states (LDOS) calculated for the main expected structures. 

As all the spectra are very different, so their control may help to distinguish grown structures.

\subsection{Vacuum Deposition}
\subsubsection{Annealing of films deposited at 60 \textdegree C}

\begin{figure}%[h]
\includegraphics[width=4cm]{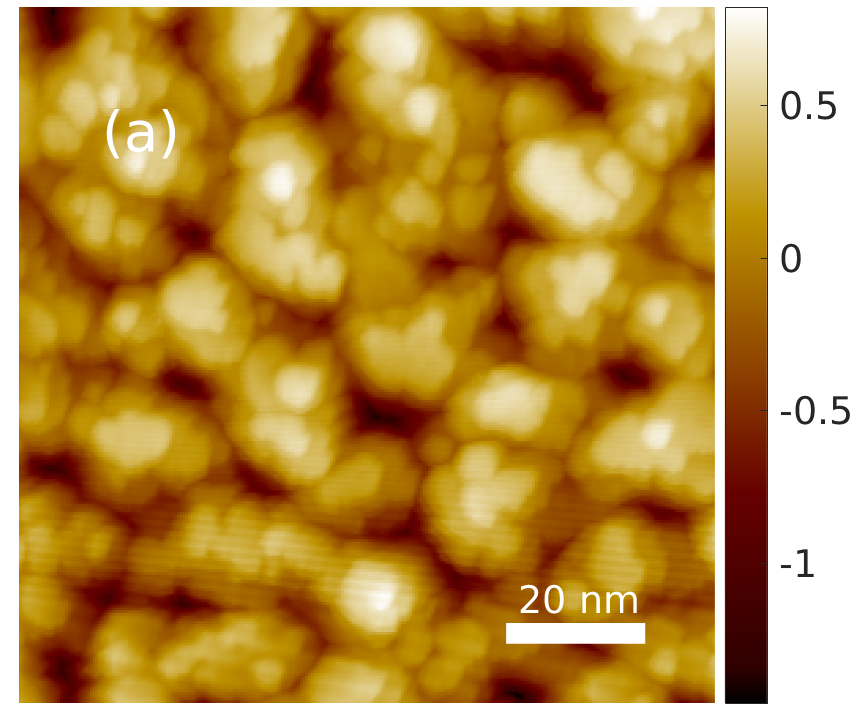}
\includegraphics[width=4cm]{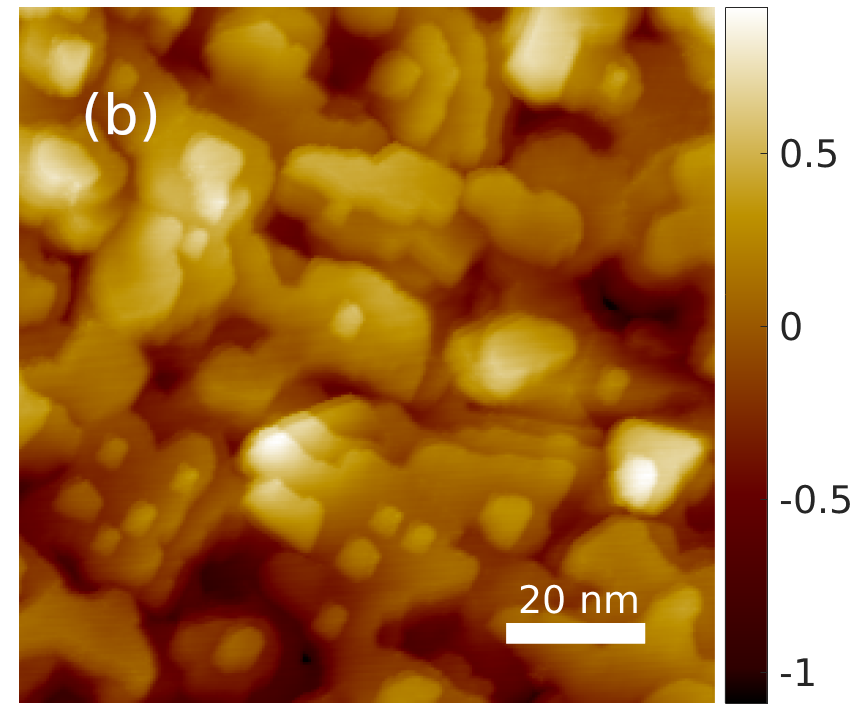}

\includegraphics[width=4cm]{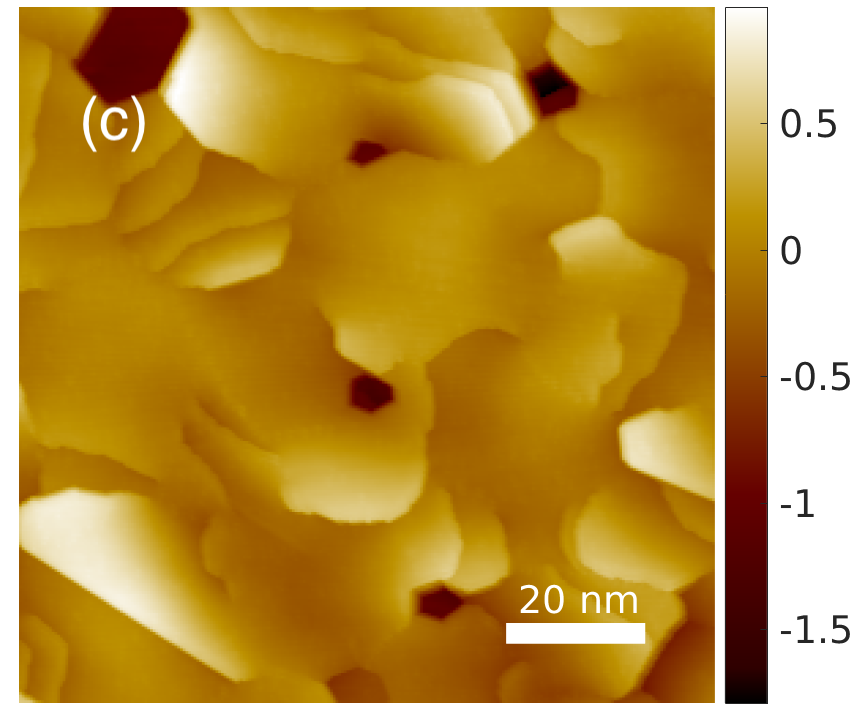}
\includegraphics[width=4cm]{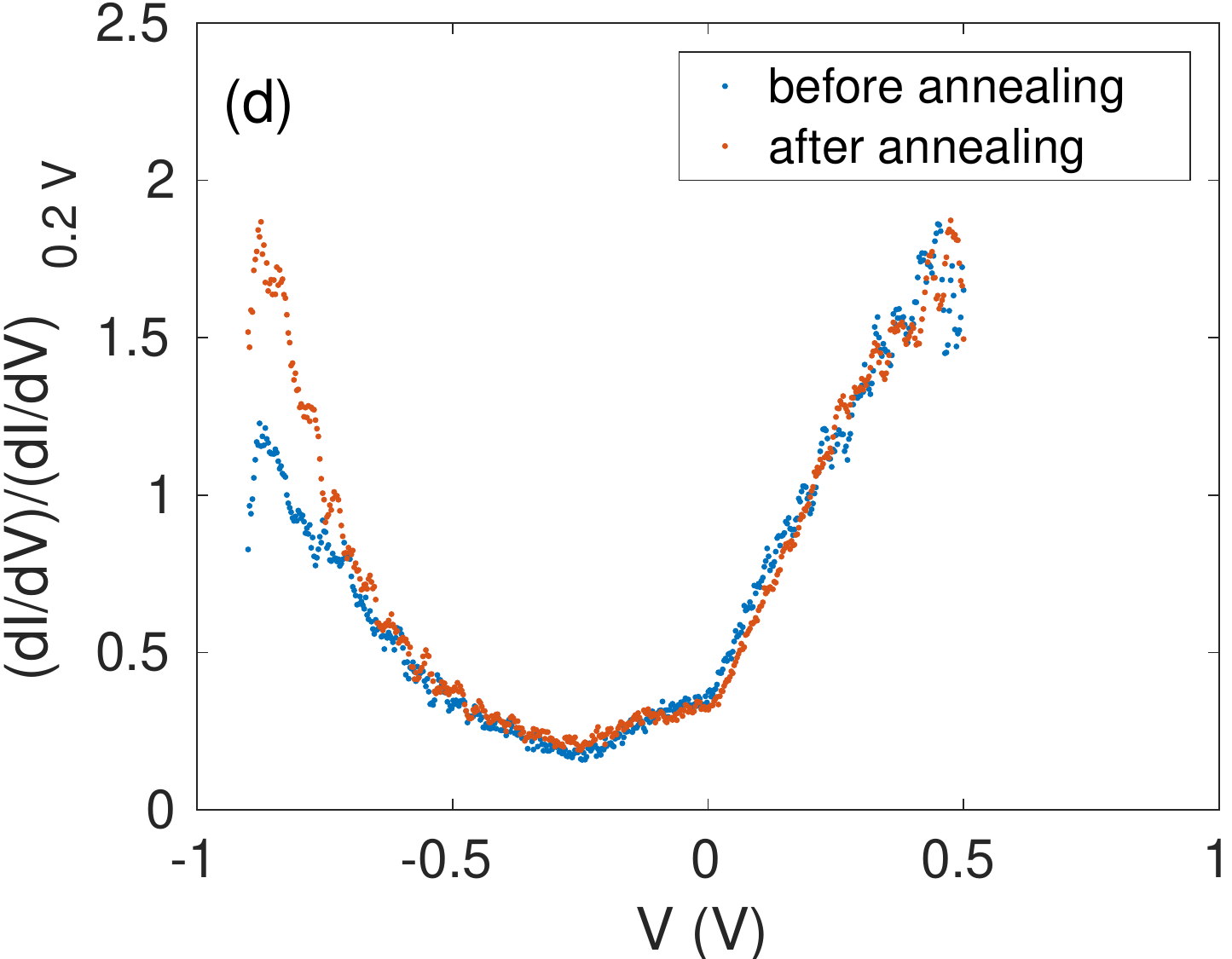}
\caption{\label{fig:50}STM images of a film deposited at substrate temperature 60 \textdegree C (a) and annealed 1 h at (b) $T_a = 150$ \textdegree C, (c) 250 \textdegree C (color scales are in nanometers). (d) Comparison of STS spectra of as-deposited and annealed at 250 \textdegree C films.}
\end{figure}

EDS analysis of films deposited at 60 \textdegree C gives ${\rm Te}/({\rm Bi}+{\rm Te})=0.65$ for relative tellurium content, that is slightly above the required 0.6 atomic percent ratio for Bi$_2$Te$_3$. %As the first step, the effect of sequential 1 hour annealing  at temperatures of 100, 150, 200, 250, 300, and 350\textdegree C on morphology of films deposited at a substrate temperature of 50\textdegree C was studied. 
Fig.~\ref{fig:50}(a) shows an STM image of a film deposited at 60 \textdegree C. The film consists of small terraced crystallites with typical sizes 10-20 nm. The characteristic terraces height on  top of the grains is 0.3-0.4 nm, which is significantly less than the thickness of the Bi$_2$Te$_3$ quintuple layer.

Heating this film to 100 \textdegree C for an hour does not lead to significant changes in its topography. However, one hour annealing at 150 \textdegree C results in a noticeable growth of lateral grain sizes (Fig.~\ref{fig:50} (b)), and grains merge into a continuous film upon further annealing at 250 \textdegree C (Fig.~\ref{fig:50} (c)). All the structures obtained with this film in the annealing process at temperatures not exceeding 250 \textdegree C look flatter and more interconnected than for films deposited at higher substrate temperature (see below). Surprisingly, the STS spectra of as-deposited and annealed films remain unchanged (Fig.~\ref{fig:50}(d)) and resemble the STS spectrum of bulk Bi$_2$Te$_3$ (Fig.~\ref{fig:STS}(f)).

An increase in the annealing temperature to 300 \textdegree C leads to the disintegration of the continuous film into separate islands 6-10 nm in height (see also results for higher deposition temperatures). 

\subsubsection{Annealing of films deposited at 175~\textdegree C and 240~\textdegree C}

\begin{figure}[h]
\includegraphics[width=4cm]{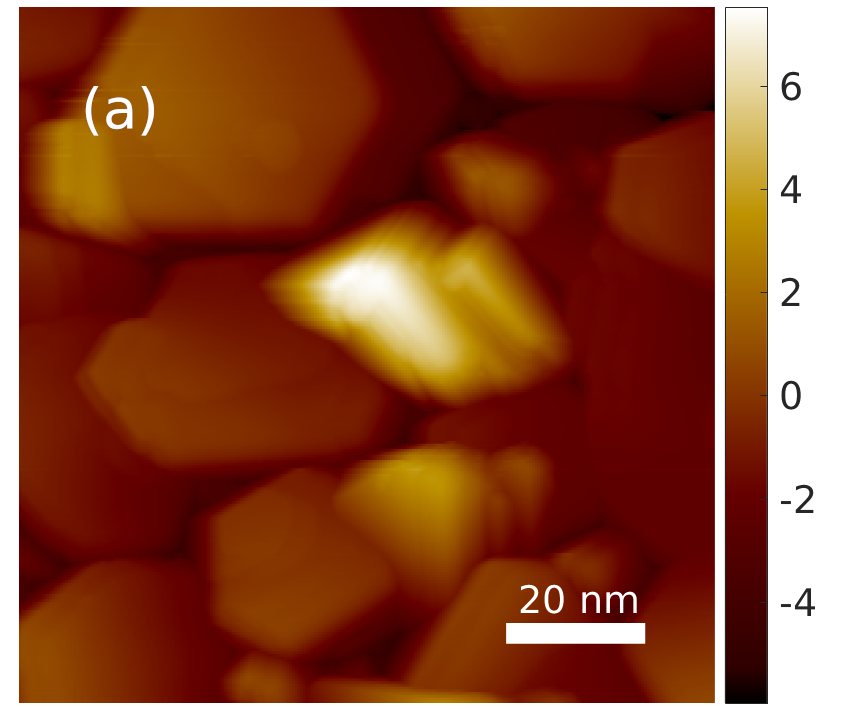}
\includegraphics[width=4cm]{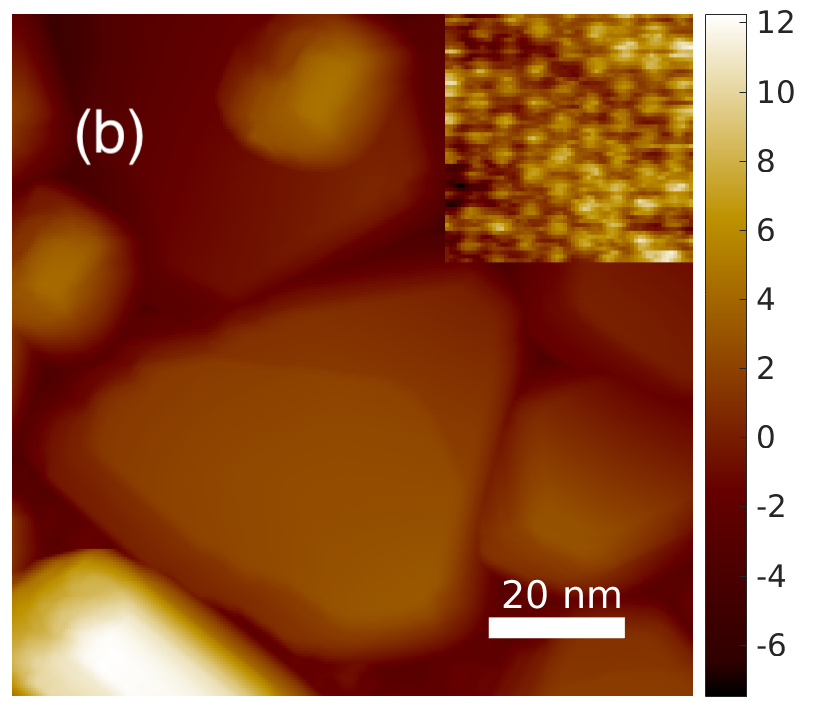}

\includegraphics[width=4cm]{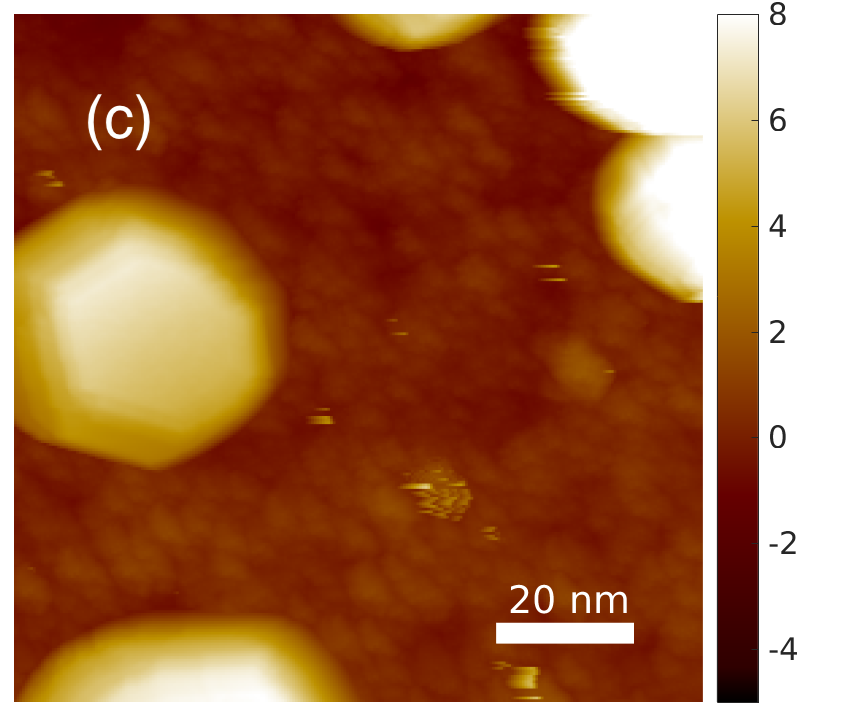}
\includegraphics[width=4cm]{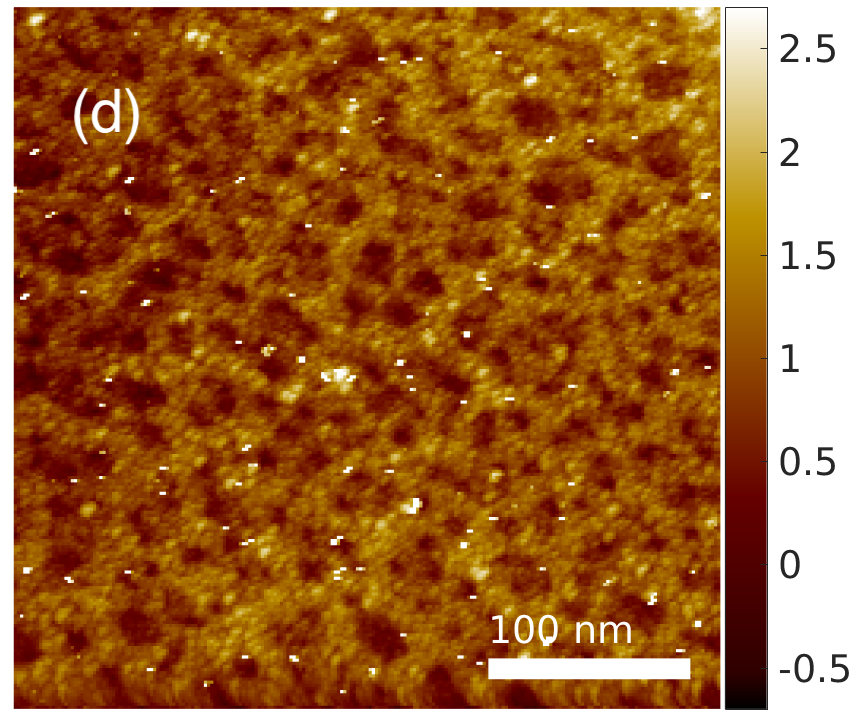}
\caption{\label{fig:200}(a) STM image of a film deposited in vacuum at substrate temperature $T_d=240$ \textdegree C. STM images of the film after 1 hour annealing at $T_a=250$ \textdegree C (b) (inset shows typical image of crystallite surface with atomic resolution at tip current 0.2 nA and sample bias -0.5 V), $T_a=300$ \textdegree C (c), and $T_a=350$ \textdegree C (d). Color scales are in nanometers.}
\end{figure}

Films deposited at  175 \textdegree C and 240 \textdegree C consist of individual grains with different orientations (Fig.~\ref{fig:200}(a)). Their faceting corresponds to the presence of symmetry axes of the 3rd order. Terraces with heights of 0.3-0.4 nm are usually observed on the upper surface of the grains. The films deposited at 240 \textdegree C have a more ordered structure and larger crystallites than films deposited at 175 \textdegree C after annealing for the same duration at the same temperatures.   In addition, the
crystallites are more disordered in films deposited at a substrate temperature of 175 \textdegree C, than in films deposited in a vacuum at 240 \textdegree C.

Annealing of the films at temperatures no higher than 250 \textdegree C leads to a gradual increase in grain sizes Fig.~\ref{fig:200}(a,b) with hexagonal symmetry of the top surface (inset in Fig.~\ref{fig:200}(b)).
Heating the film to 300 \textdegree C for one hour changes the picture - the grains become solitary -
 and also opens the surface of the substrate, which is a disordered layer (Fig.~\ref{fig:200}(c)). Annealing at $T = 350$ \textdegree C results in disappearance of the grains but not of the disordered layer (Fig.~\ref{fig:200}(d)).

The grains are 6-8 nm in height and
the atomic structure on their tops remains almost unchanged and has hexagonal symmetry, in accordance with observed grain faceting. In general, the observed behavior corresponds to
growth of Te-deficient structures such as (Bi$_2)_n$/(Bi$_2$Te$_3$)$_m$ ($n>0$).

To summarize, PLD under vacuum conditions allows us to grow polycrystalline films of Bi$_2$Te$_3$ with grain sizes varying from tens to few hundreds of nanometers. Subsequent annealing in the temperature region 150-350\textdegree C leads to growth of lateral grain sizes, then formation of sets of independent grains and finally to complete sublimation of crystallites leaving a disordered layer. The presence of this layer and varying orientations of the crystallites  indicate chemical interaction of the film with the substrate surface and thus the absence of  vdW epitaxy. We attribute this fact to 
the damage by the  high-energy ions produced in the process of the laser ablation of the target. Their energies may reach hundreds eV, as was reported in \cite{Franghiadakis_1999}. Such high-energy ions would damage the substrate, thereby impairing the vdW epitaxy capability.

\subsection{Deposition in argon}

 A way to avoid direct ion bombardment of a substrate is to perform PLD in the presence of an inert background gas. The effect of argon background gas pressure on the results of PLD of Bi$_2$Te$_3$ was reported earlier \cite{Bailini2007}. The best quality of films  was achieved at an argon pressure of 40 Pa (0.3 Torr). So  we also used argon at a pressure of 0.3 Torr as a background gas for the present study. As the room-temperature mean-free path is below 1 mm at this pressure, such a pressure is enough to exclude completely the  high-energy ion damage of the substrate.

 Fig.~\ref{fig:EDS_Ar} shows the atomic composition of thick films (typical nominal thickness  100 nm) deposited in argon at different substrate temperatures. The films deposited at $T\lesssim 165$ \textdegree C show very little Te loss with respect to the composition of the target, that is they contain a large excess amount of Te. At the same time, the films deposited at $T\gtrsim 220$ \textdegree C exhibit a slight Te deficiency. These data are in agreement with our preliminary Te deposition experiments.

\subsubsection{Annealing of films deposited at 50 \textdegree C}

\begin{figure}%[h]
\center{\includegraphics[width=7cm]{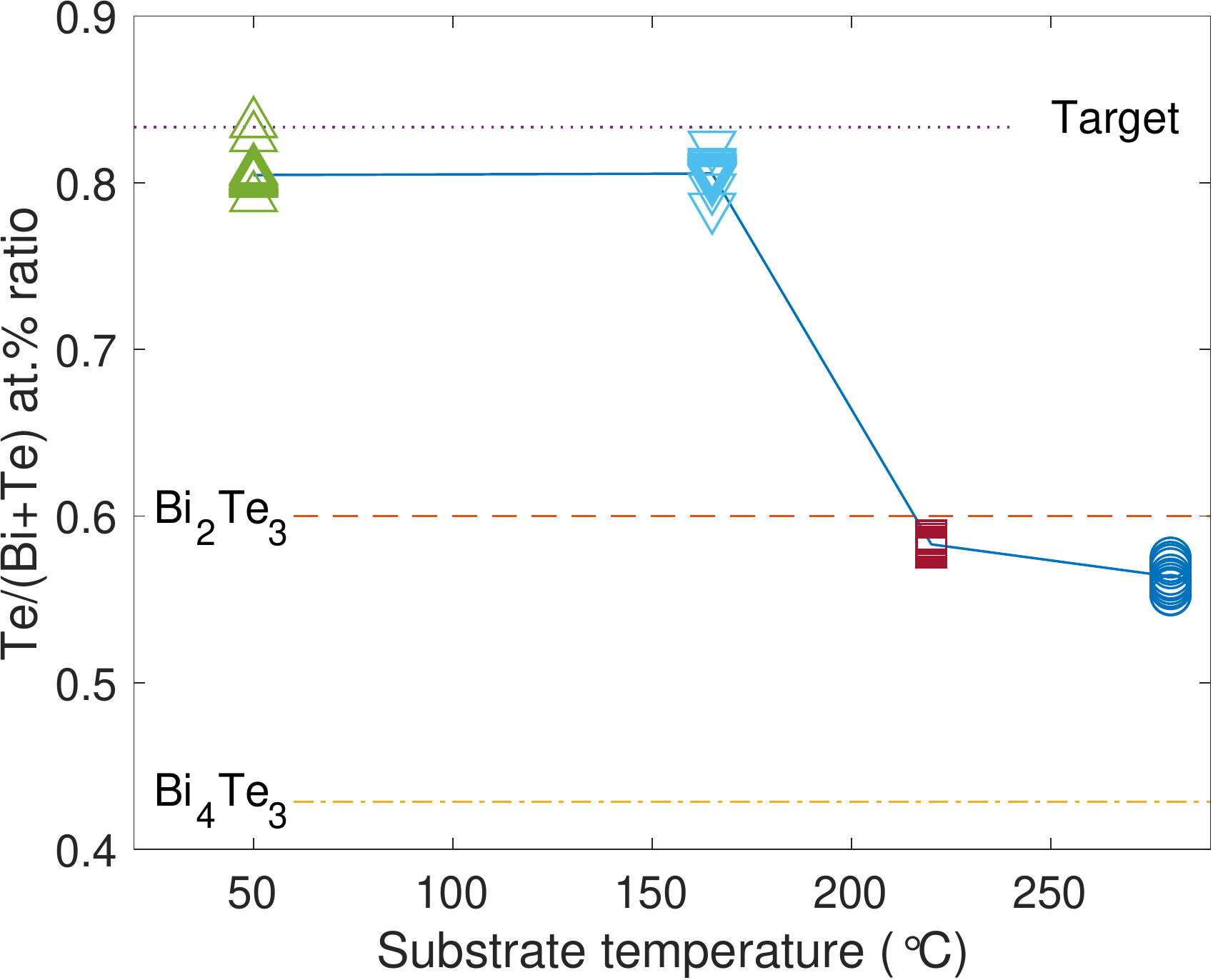}}

\caption{\label{fig:EDS_Ar} Atomic composition of films deposited by PLD under 0.3 Torr of argon at different temperatures.}
\end{figure}

\begin{figure}%[h]
\includegraphics[width=4.1cm]{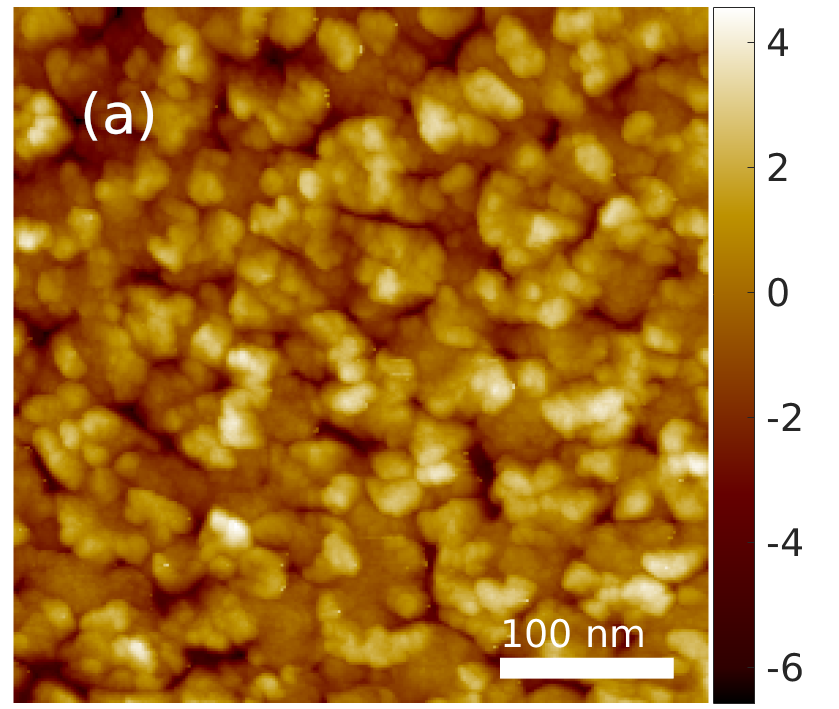}
\includegraphics[width=4.1cm]{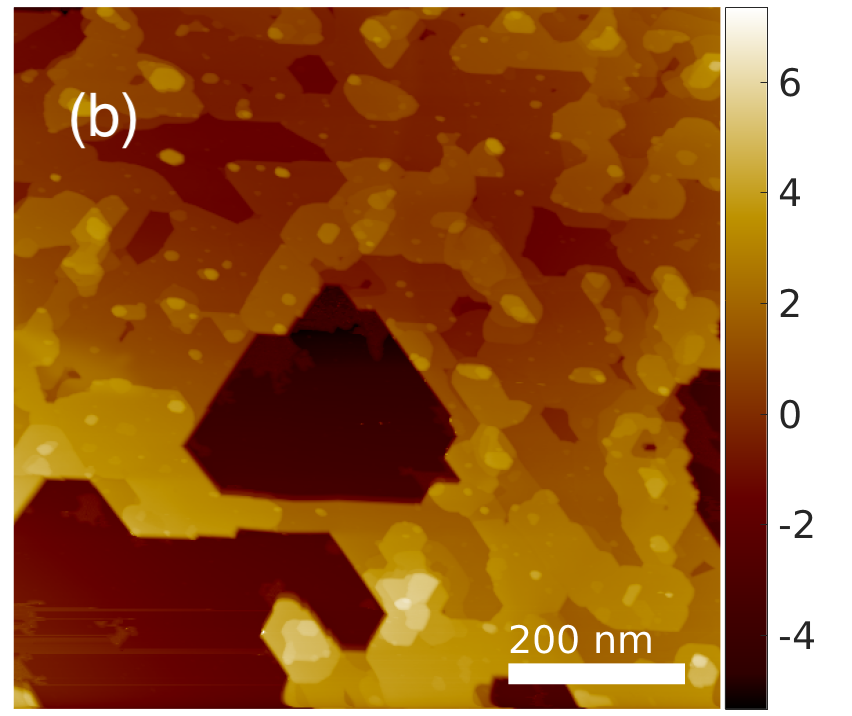}

\includegraphics[width=4.1cm]{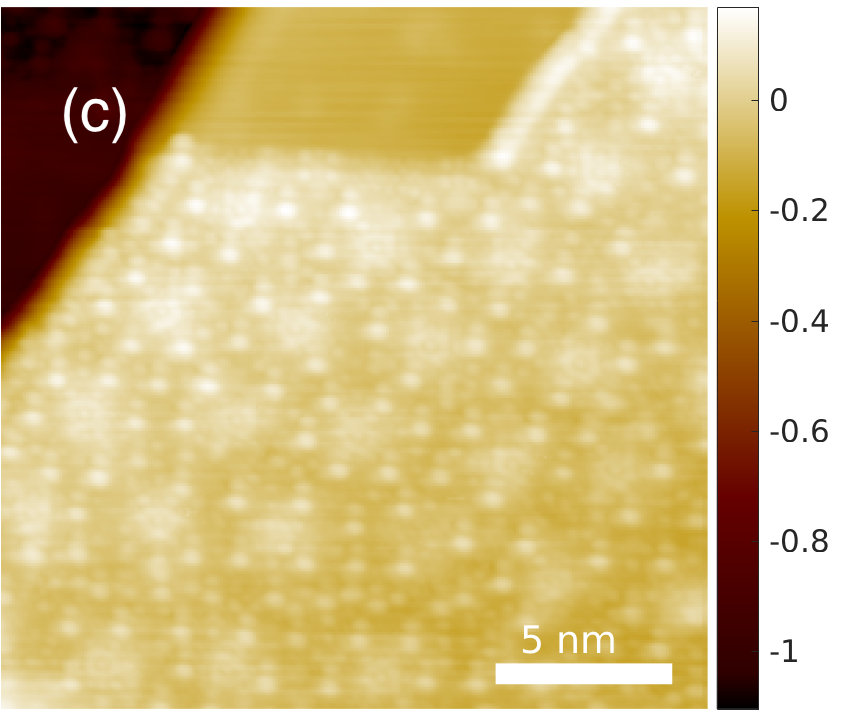}
\includegraphics[width=4.1cm]{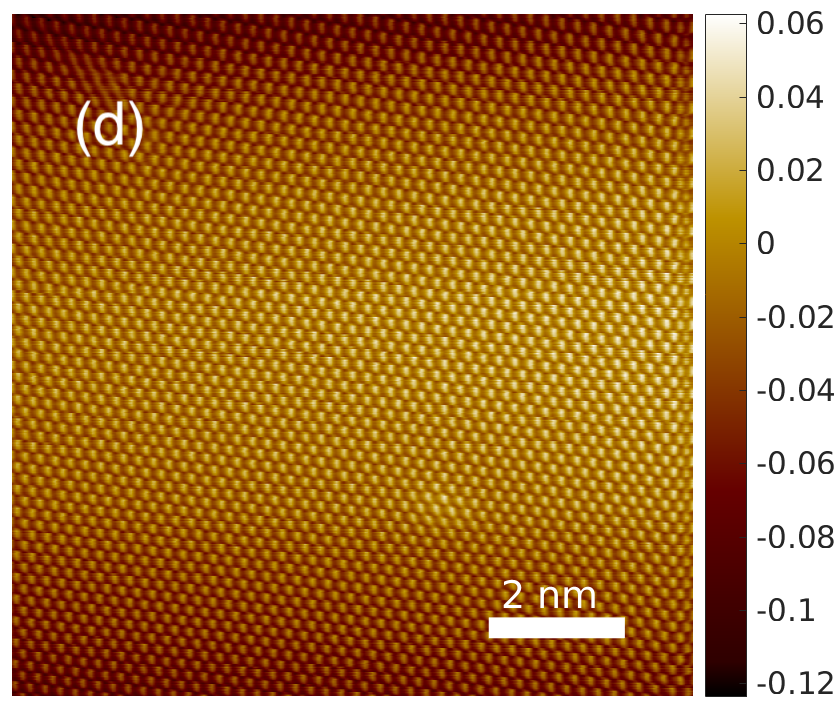}

\caption{\label{fig:a50}(a) STM image of a film deposited at substrate temperature 50 \textdegree C at argon pressure 0.3 Torr; the same film after annealing at $T=300$ \textdegree C (b,c), and $T=350$ \textdegree C (d) (color scales are in nanometers).}
\end{figure}

Films deposited at 50 \textdegree C consist of crystallites 5-10 nm in lateral sizes (Fig. \ref{fig:a50}(a)). The EDS analysis gives the presence of excess tellurium with ${\rm Te/Bi} = 4.1\pm 0.05$ atomic percent ratio (Fig.~\ref{fig:EDS_Ar}). Subsequent annealing for 1 hour at  100~\textdegree C leads to a slight increase in the size of crystallites. Further annealing for 1 hour at 150~\textdegree C does not fundamentally change the picture. Crystallites with the atomic structures characteristic of Bi$_2$Te$_3$  as well as Te can be observed. 

The morphology of the film remains almost the same for $T_a\lesssim 200$~K and starts to change significantly after annealing at $T_a = 250$ \textdegree C and above. After annealing at $T_a=250$ \textdegree C, the crystallites acquire a finished faceting, and for $T_a =300$ \textdegree C, the film 
becomes flat and continuous. It also partially evaporates, uncovering intact HOPG surface. The terrace edges become straight, and faceted holes in the films develop (Fig. \ref{fig:a50}(b)).
In this case, the orientation of the boundaries of the terraces corresponds to the orientation of the graphite substrate. Thus, the graphite substrate has an orienting effect on the film, despite the mismatch between the lattices of the substrate and the film. The surface of the films annealed at 300 \textdegree C  shows indications of partial film decomposition, namely spots of rough surfaces with very interesting atomic ordering (Fig. \ref{fig:a50}(c)). The origin of such an ordering remains unclear and may be a manifestation of  charge-density waves (CDW) with the wave vector 1.6 nm$^{-1}$ accompanied by a surface reconstruction.  Further heating up to 350 \textdegree C leads to complete evaporation of the films, leaving a clean substrate surface (Fig. \ref{fig:a50}(d)). Thus, the use of background gas during PLD not only promotes more adequate transfer of the target material to the substrate, but also protects the original structure of the substrate.

\subsubsection{Annealing of films deposited at 165 \textdegree C}
\begin{figure}%[ht]
\includegraphics[width=4.1cm]{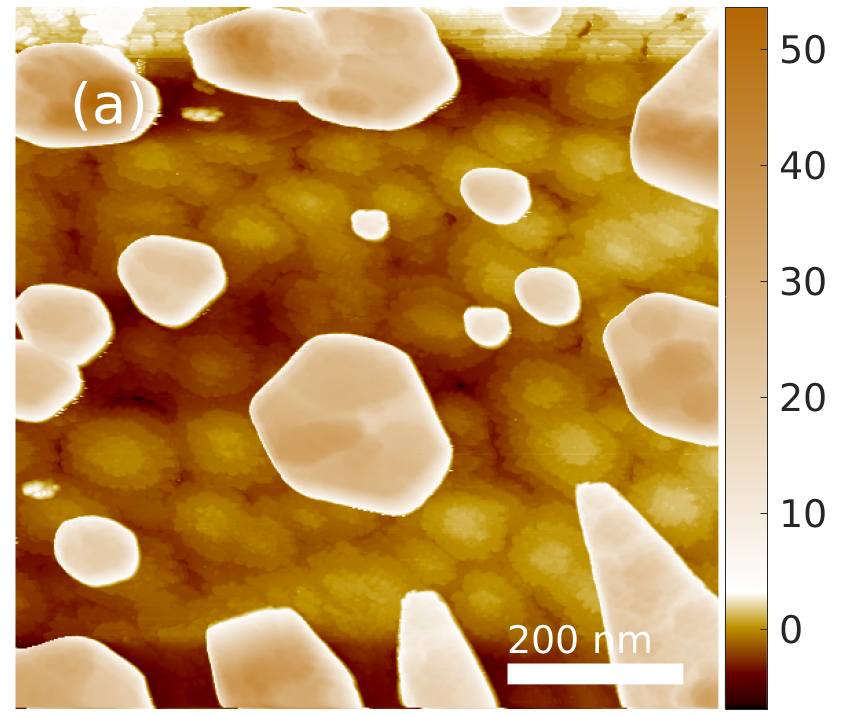}
\includegraphics[width=4.1cm]{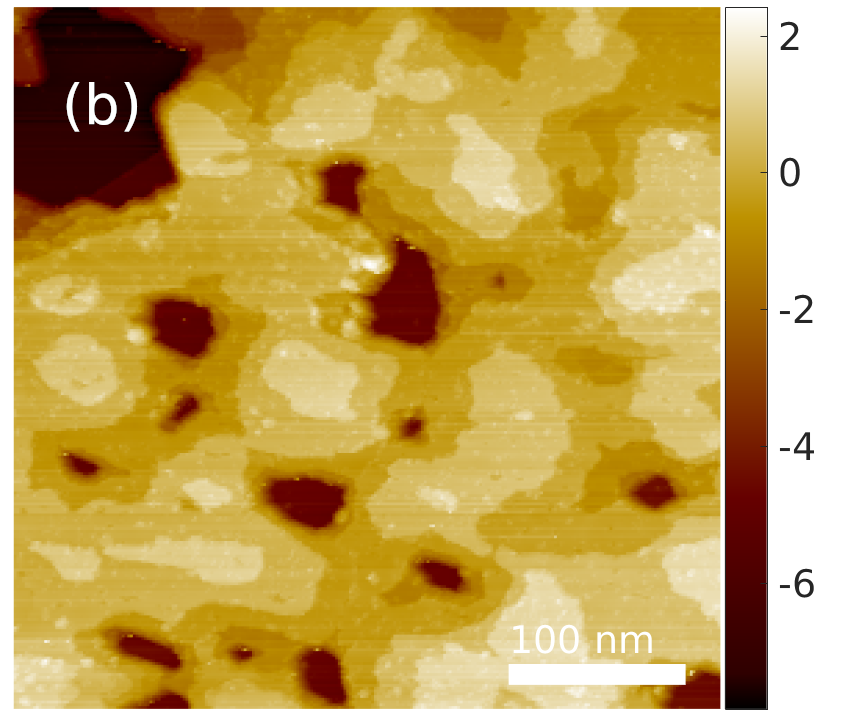}

\includegraphics[width=4.1cm]{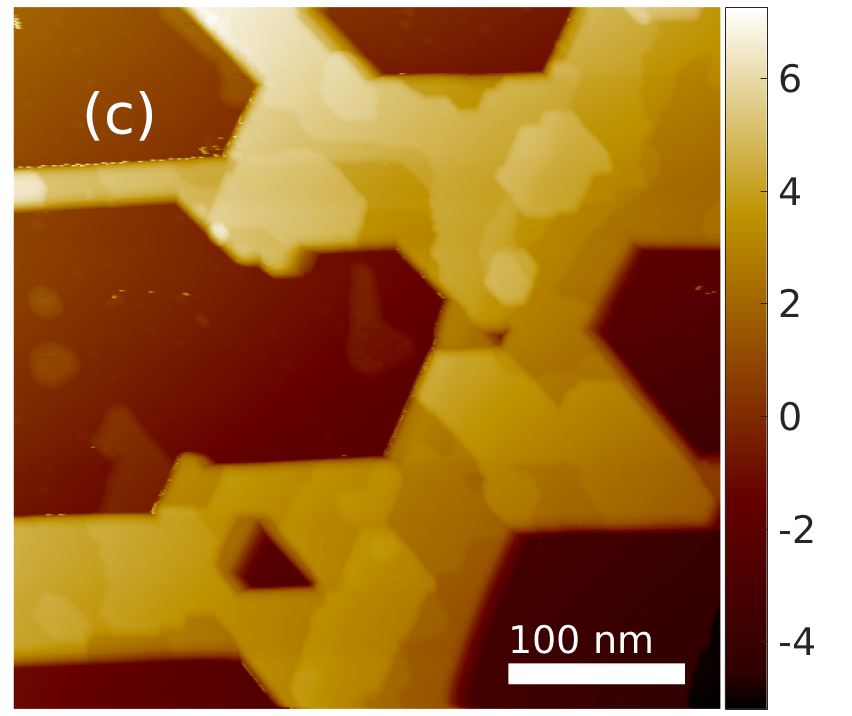}
\includegraphics[width=4.1cm]{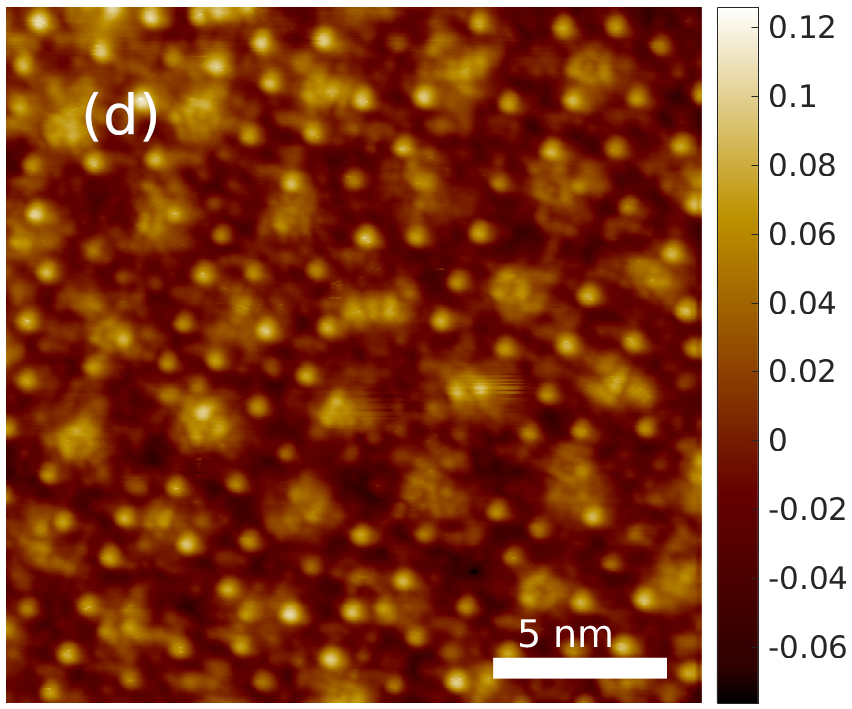}
\caption{\label{fig:a150}(a) STM image of a film deposited at substrate temperature 165 \textdegree C at argon pressure 0.3 Torr;   The same film after 1 hour annealing at (b) 250 \textdegree C, and (c) 300 \textdegree C; (d) the upper surface of the film after annealing at 300 \textdegree C. Color scales are in nanometers.}
\end{figure}

Films deposited at 165 \textdegree C in argon atmosphere demonstrate  formation of large crystallites with well defined faces  in addition to a continuous film (Fig.~\ref{fig:a150}(a)). Their typical heights are few tens of nanometers and correlate with their cross-section area in agreement with the Gibbs-Wulff theorem. The crystallites have the same orientation of faces. We attribute their formation to the excess tellurium.% and absorb excess tellurium.

The continuous film consists of small terraces that form a discrete hilly relief. The faces of terraces are not well defined and have a wavy shape. 

One hour annealing of the film at 200 \textdegree C leads to almost complete disappearance of the large crystallites and formation of  a film with $\approx 1$ nm height terraces. Further annealing at 250 \textdegree C during 1 hour leads to further growth of holes down to the substrate surface without noticeable change of the film thickness (Fig.~\ref{fig:a150}(b)). The dominating terrace height remains to be $\approx 1$ nm. 

The final 1 hour annealing at 300 \textdegree C leads to formation of a crystalline mesh (Fig.~\ref{fig:a150}(c)). The height of the mesh $\approx 5$-6~nm is still very close to the initial value of the film thickness (Fig.~\ref{fig:a150}(a,b)), but the structure of the upper surface (Fig.~\ref{fig:a150}(f)) demonstrates the same atomic reconstruction as films deposited at 50  \textdegree C and annealed at 300  \textdegree C (Fig.~\ref{fig:a50}(c)).

\begin{figure}%[ht]
\includegraphics[width=4.1cm]{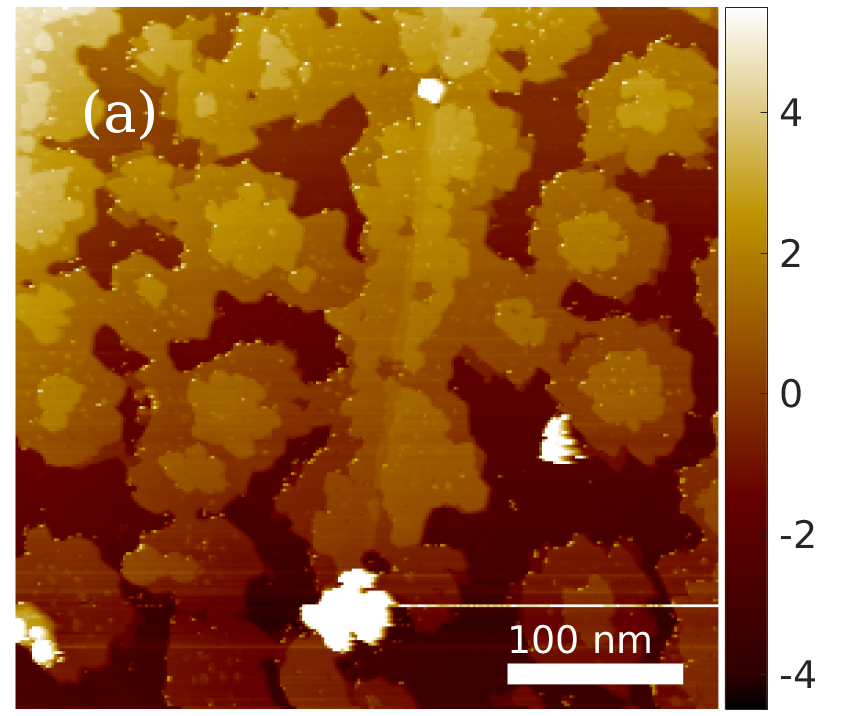}
\includegraphics[width=4.1cm]{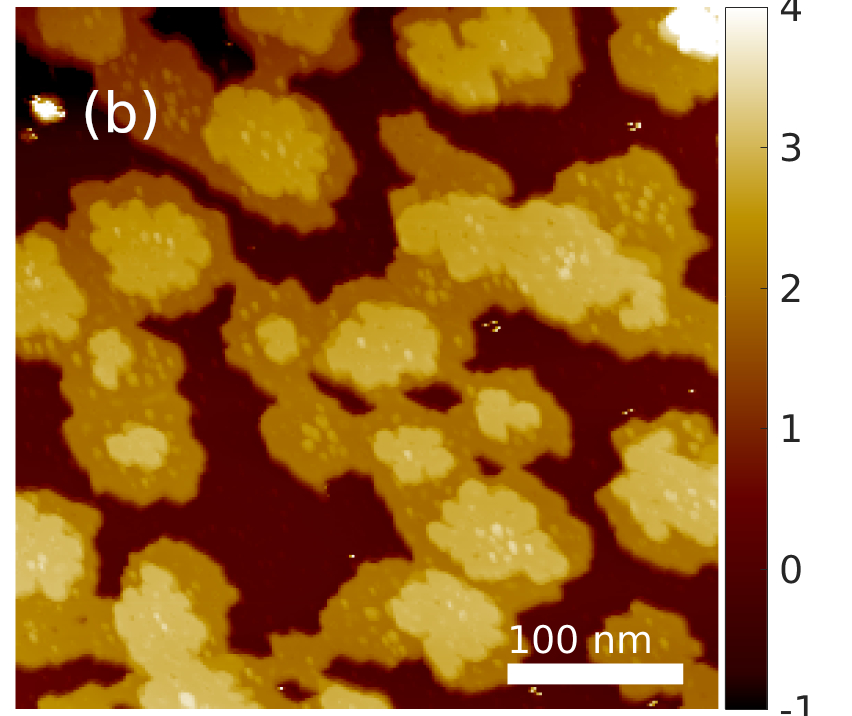}

\includegraphics[width=4.1cm]{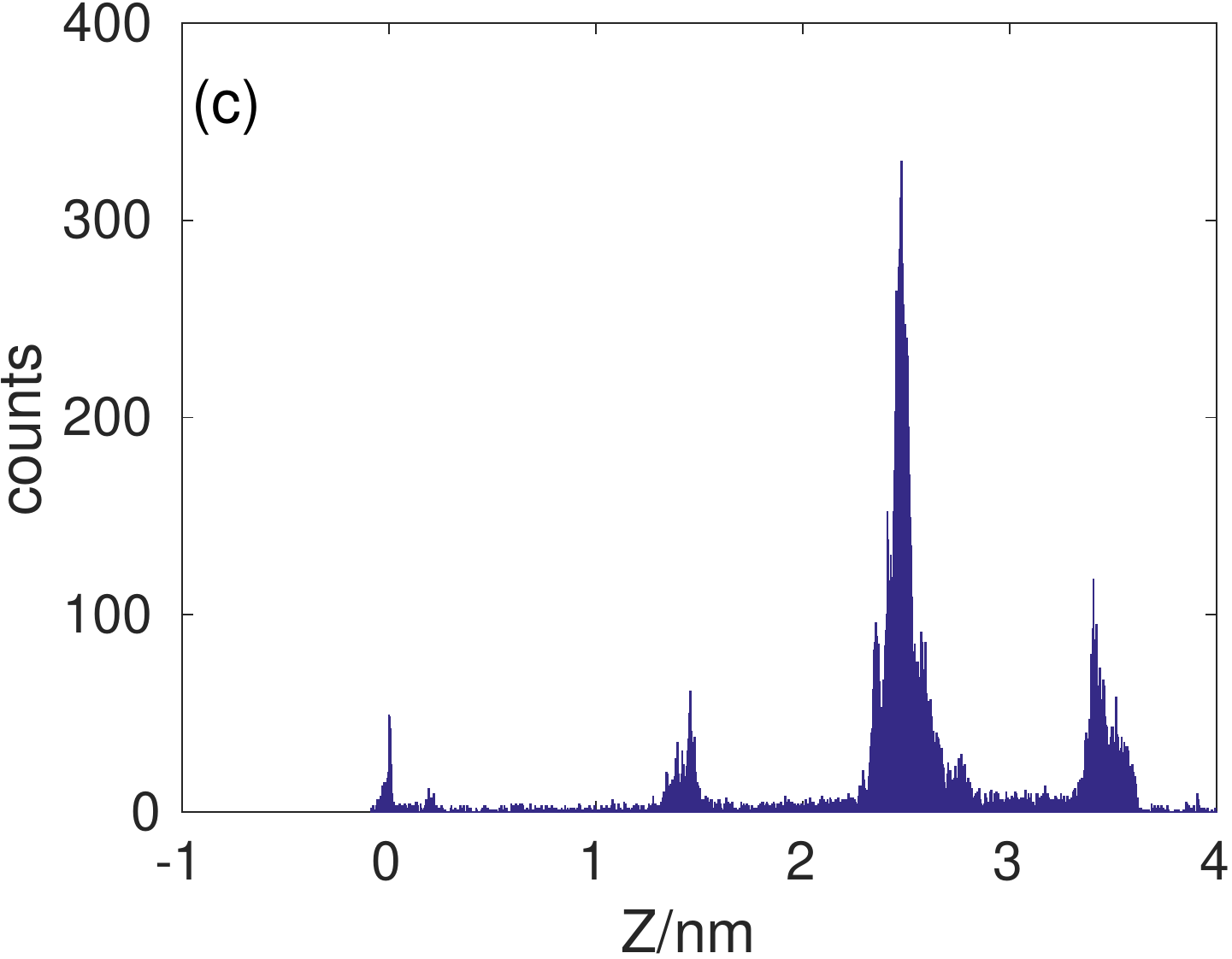}
\includegraphics[width=4.1cm]{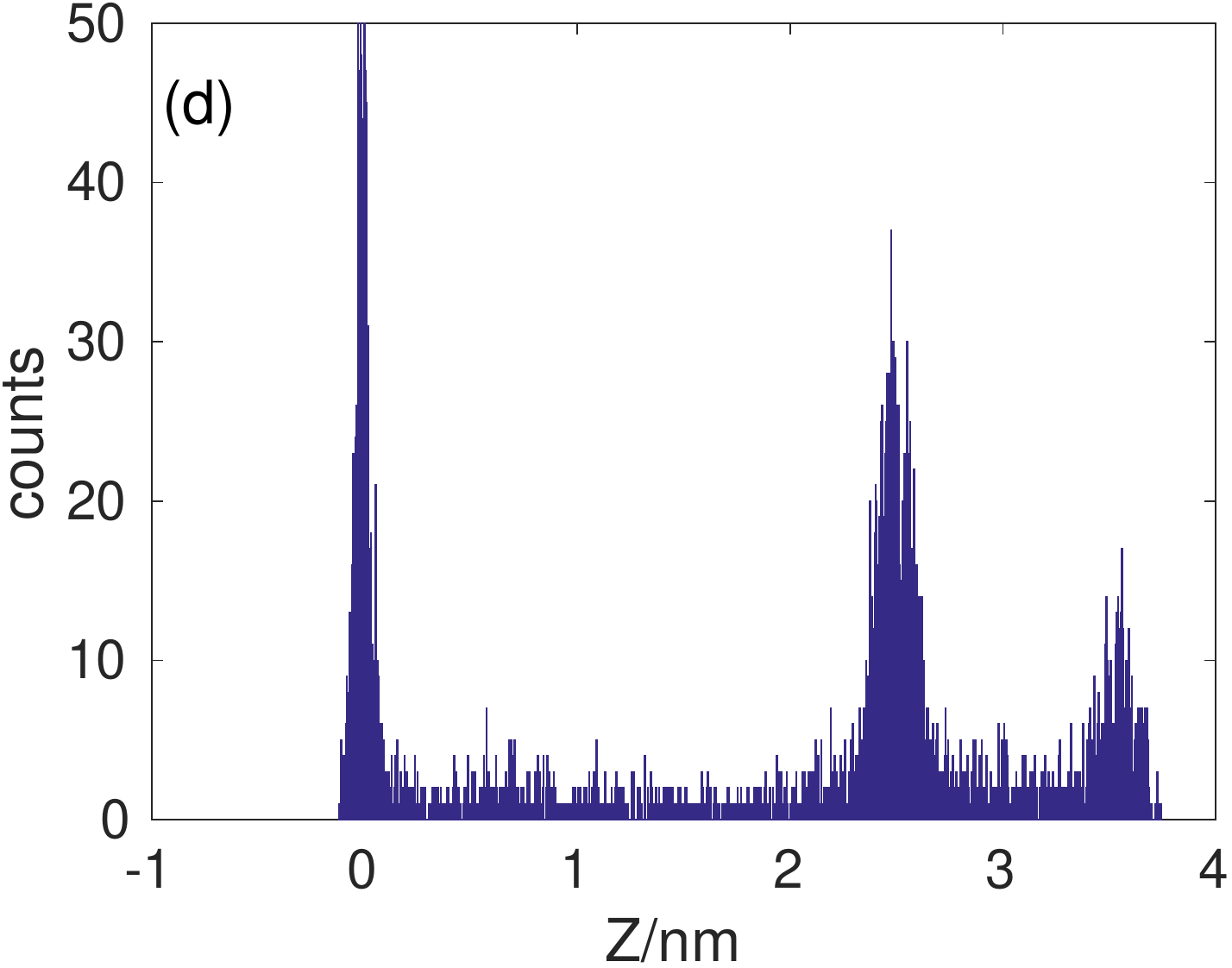}

\includegraphics[width=4.1cm]{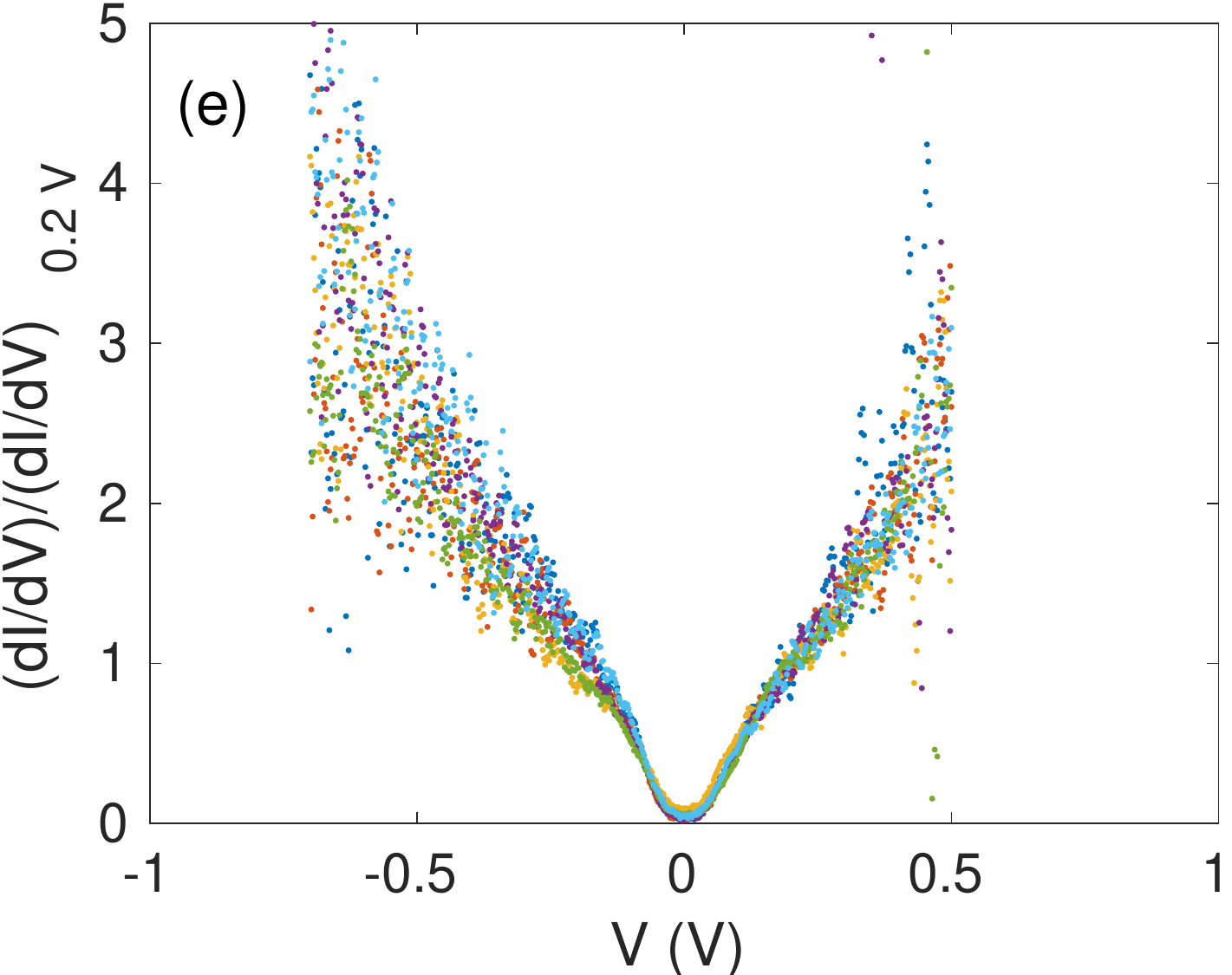}
\includegraphics[width=4.1cm]{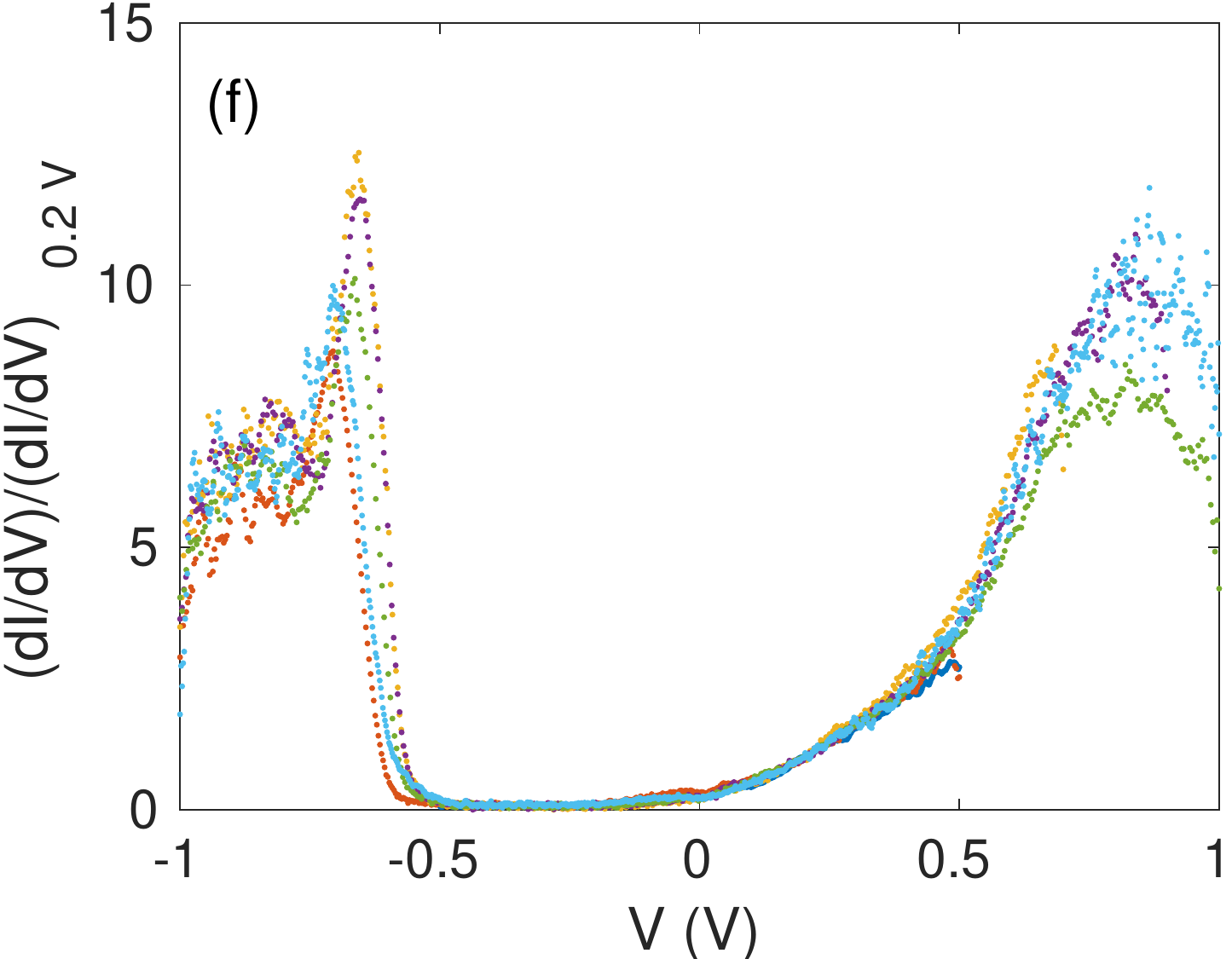}

\includegraphics[width=4.1cm]{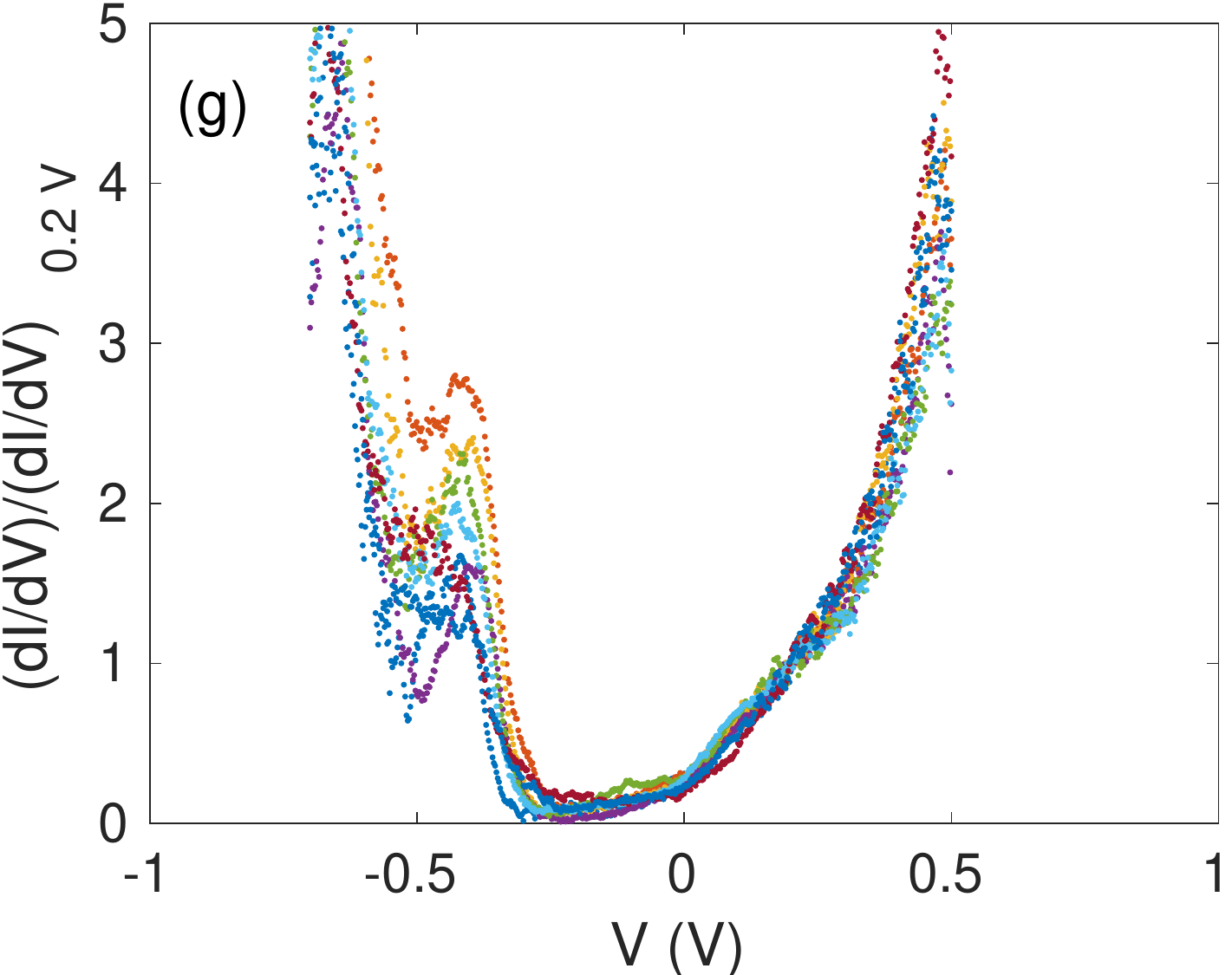}
\includegraphics[width=4.1cm]{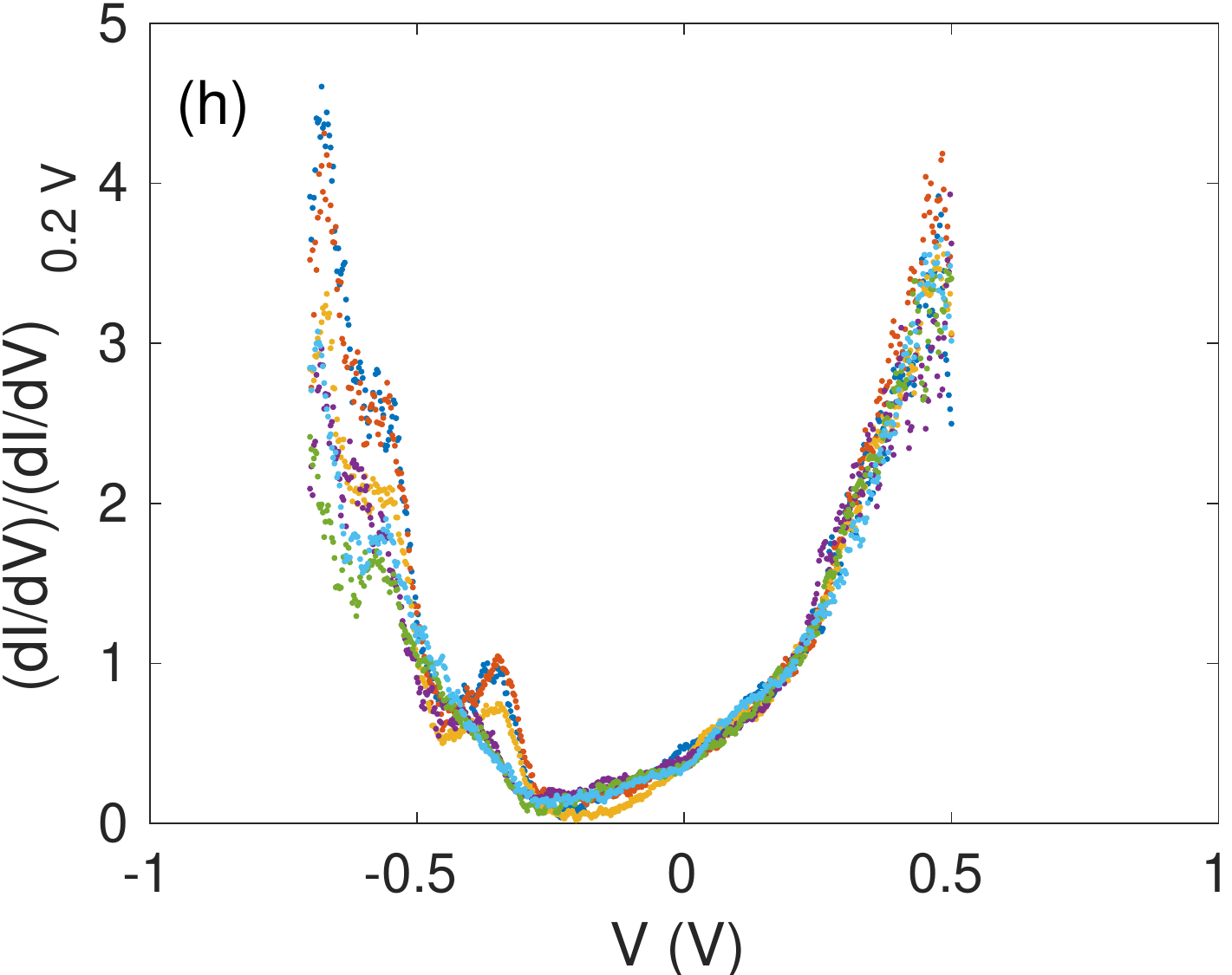}

\caption{\label{fig:a150sts} STM images of a thin film  deposited at 165 \textdegree C and (a) annealed at 200 \textdegree C; (b) further annealed at 250 \textdegree C (color scales are in nanometers). (c),(d) - typical height distribution histograms of the film shown in (a), and (b) respectively.
STS spectra of ultrathin films deposited at 165 \textdegree C and annealed at 200 \textdegree C collected at different levels: (e) substrate level; (f) first terrace; (g) second terrace; (h) third terrace.}
\end{figure}

Film diagnostics is convenient to perform on thinner films consisting of a few QL. Fig.~\ref{fig:a150sts}(a,b) shows STM images of such a film. It consists of partially coalesced islands. The islands consist of three layers after annealing at 200 \textdegree C (Fig.~\ref{fig:a150sts}(a). The heights analysis (Fig.~\ref{fig:a150sts}(c)) gives SL thickness for the first one and QL thickness for the next two. Annealing at 250 \textdegree C results in merging of the first two layers (Fig.~\ref{fig:a150sts}(b,d)). STS data of the substrate, 1st, 2nd and 3rd layers are shown in Fig.~\ref{fig:a150sts}(e-h) respectively. The data demonstrate very good correspondence to the calculated spectra for 1, 2, and 3QL (Fig.~\ref{fig:STS}(d-f)) respectively.

\subsubsection{Annealing of films deposited at 220 \textdegree C}

\begin{figure}[h]
\includegraphics[width=4.1cm]{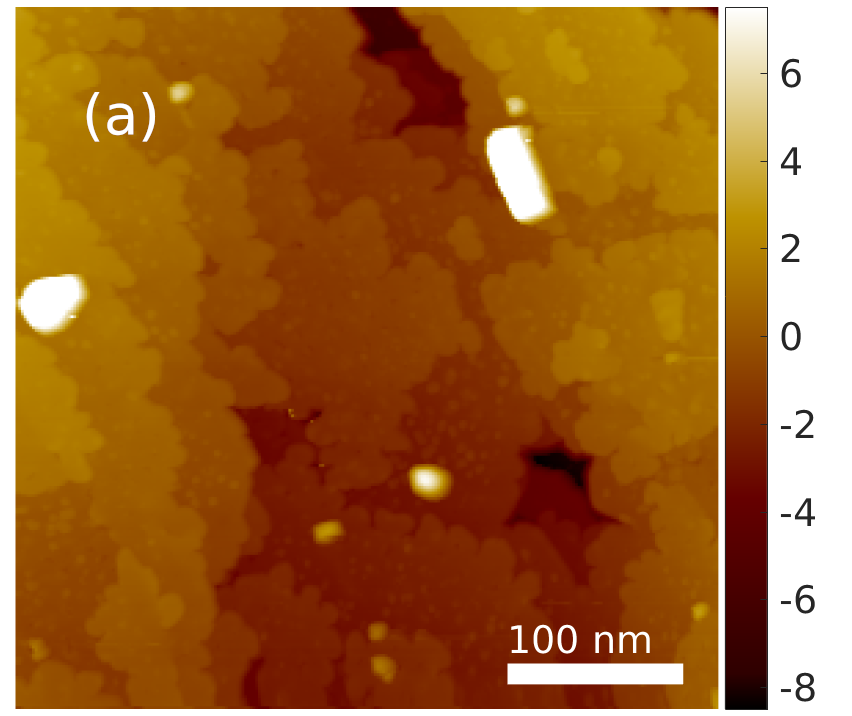}
\includegraphics[width=4.1cm]{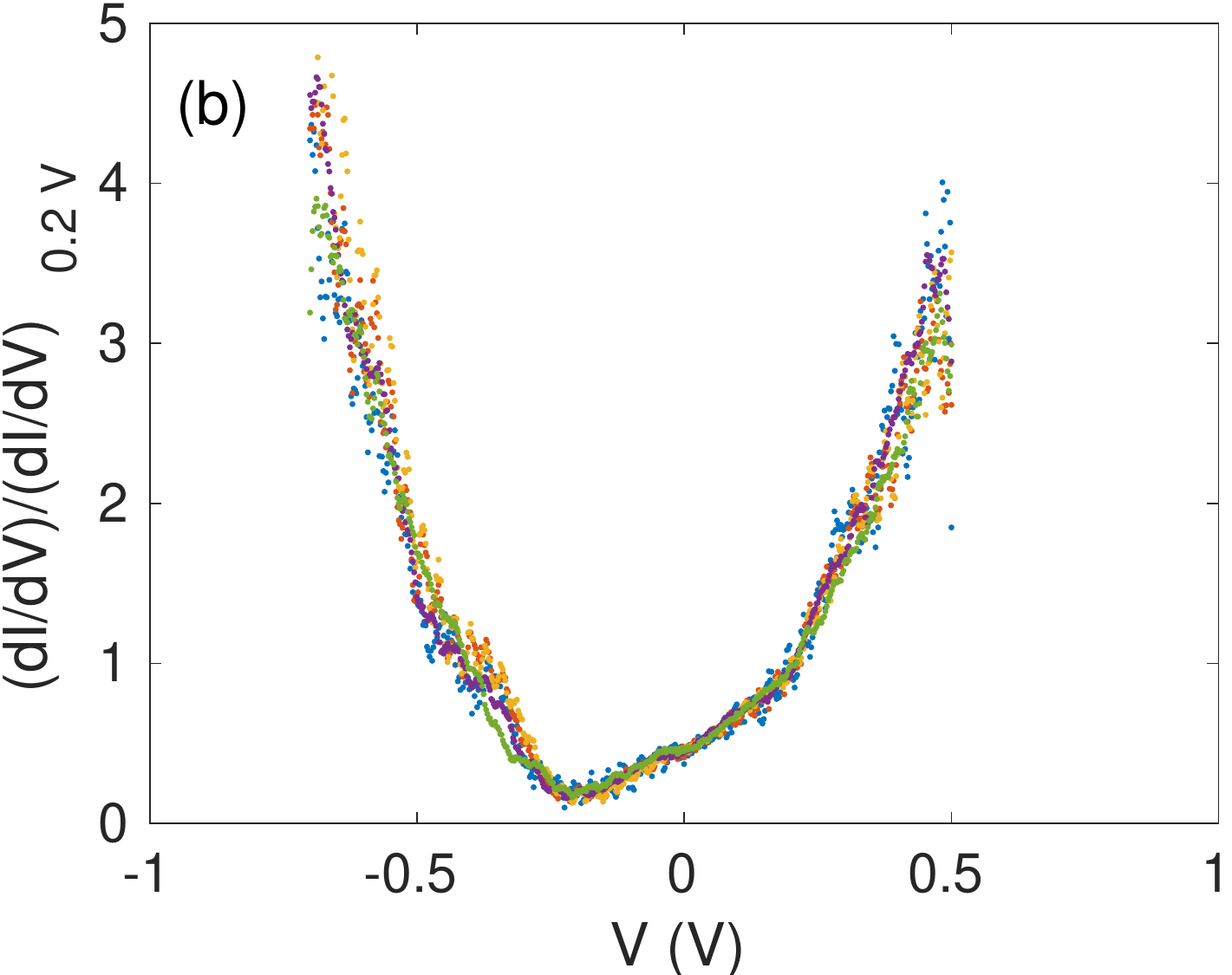} 
\caption{\label{fig:a200}
(a) STM image of a film deposited at substrate temperature 220 \textdegree C and argon pressure 0.3 Torr (color scale is in nanometers);
(b) LDOS of the film.}
\end{figure}

Fig. \ref{fig:a200}(a) shows a typical STM image of a film deposited at $T_d= 220$\textdegree C. The film consists of well-developed terraces with typical thickness 1 nm which corresponds to 1 QL of Bi$_2$Te$_3$. Tunneling LDOS of the film (Fig. \ref{fig:a200}(d)) corresponds to LDOS of Bi$_2$Te$_3$ (Fig.~\ref{fig:STS}(f)) with the chemical potential level inside the bulk band gap.  The main difference between the films deposited at $T_d=220$ \textdegree C and deposited at smaller temperature and annealed at $T > 200$ \textdegree C is that 
the latter have a tendency to grow as a set of islands, whereas the former form quasi-continuous layers.

One hour annealing of the film at  $T_a=250$ \textdegree C gives the same effect as is observed in films deposited in argon at  50 \textdegree C and annealed at 250 \textdegree C. Namely, the terrace boundaries are getting more and more flat and a hole structure down to the substrate starts to develop. 

\subsubsection{Deposition at 280 \textdegree C}

\begin{figure}[h]
\includegraphics[width=4.1cm]{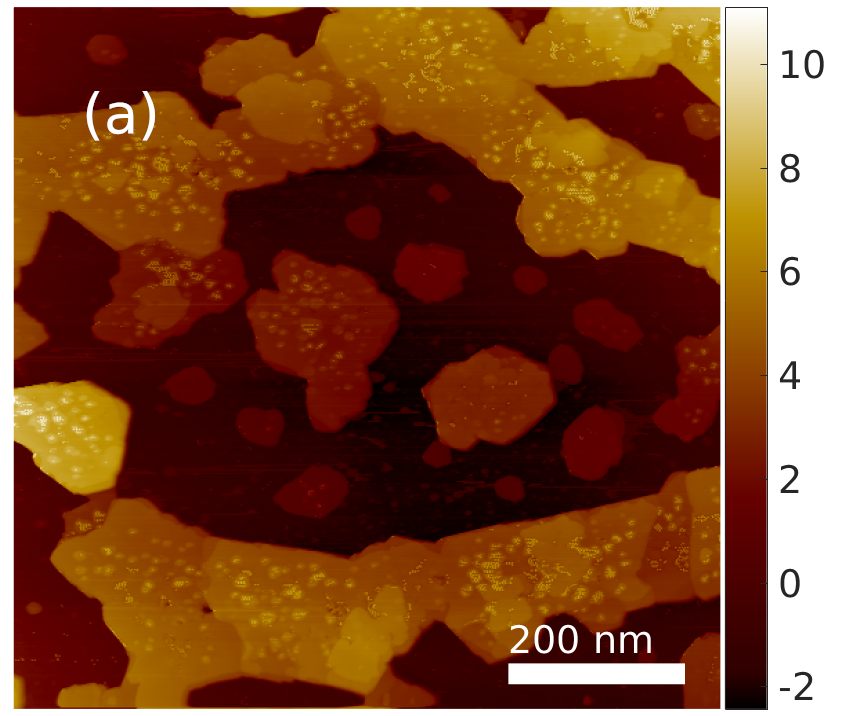}
\includegraphics[width=4.1cm]{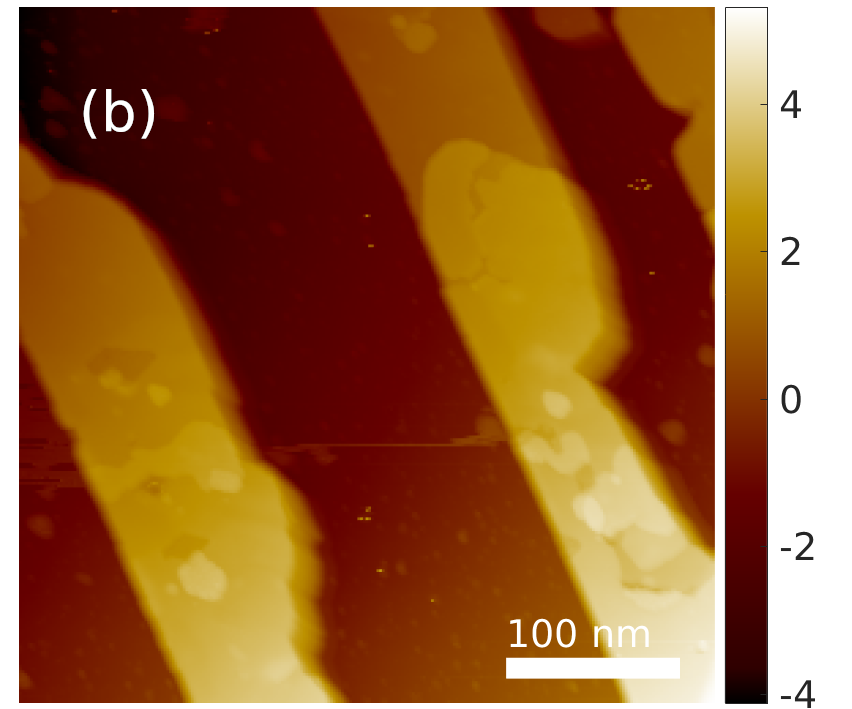}

\includegraphics[width=4.1cm]{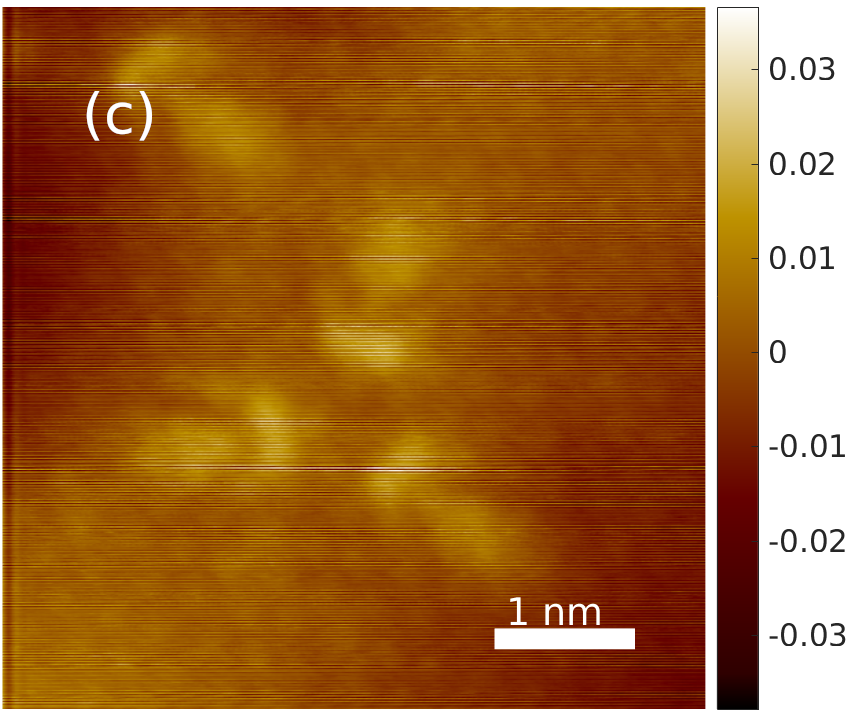}
\includegraphics[width=4.1cm]{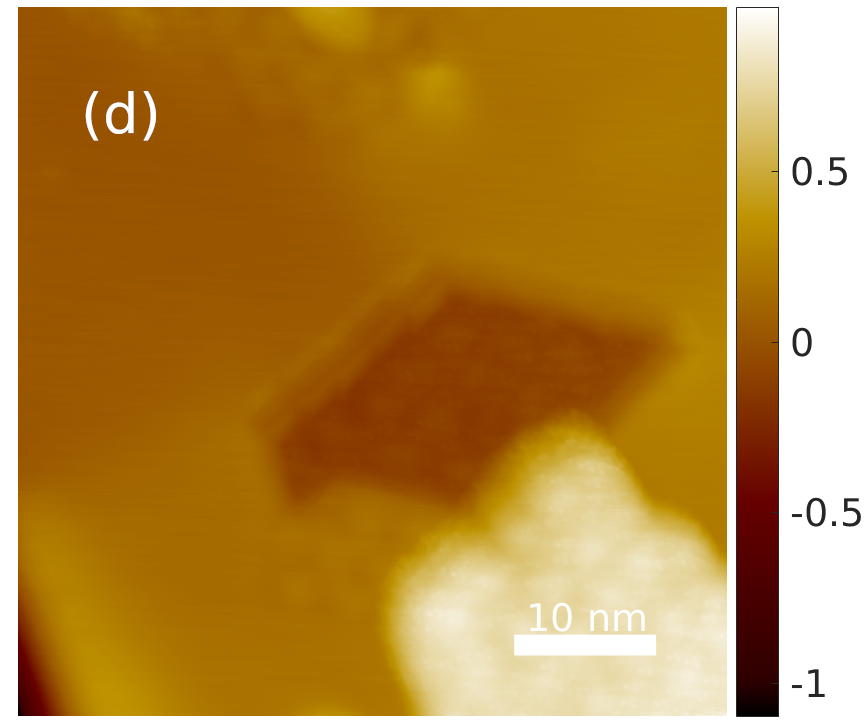}

\includegraphics[width=4.1cm]{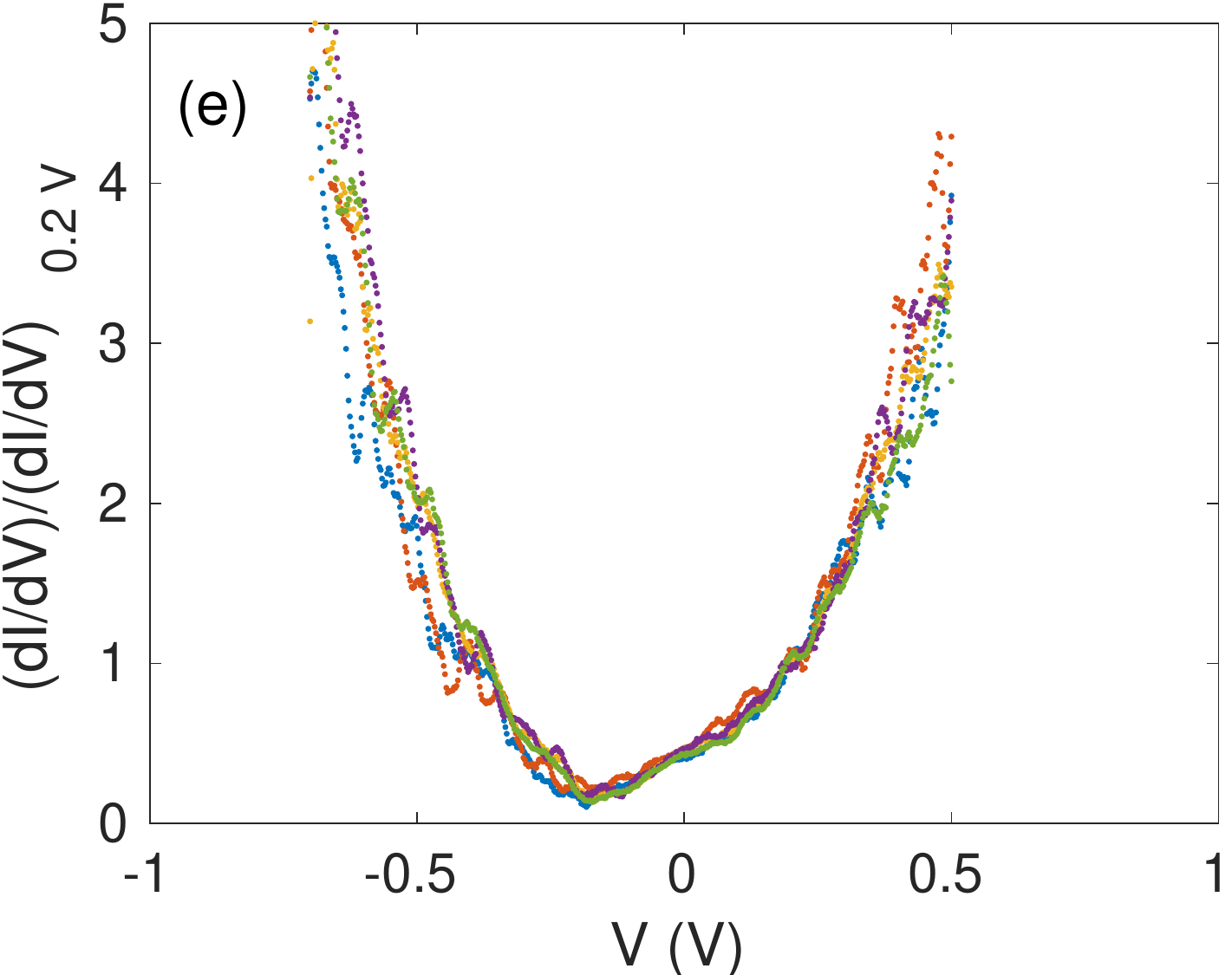}
\includegraphics[width=4.1cm]{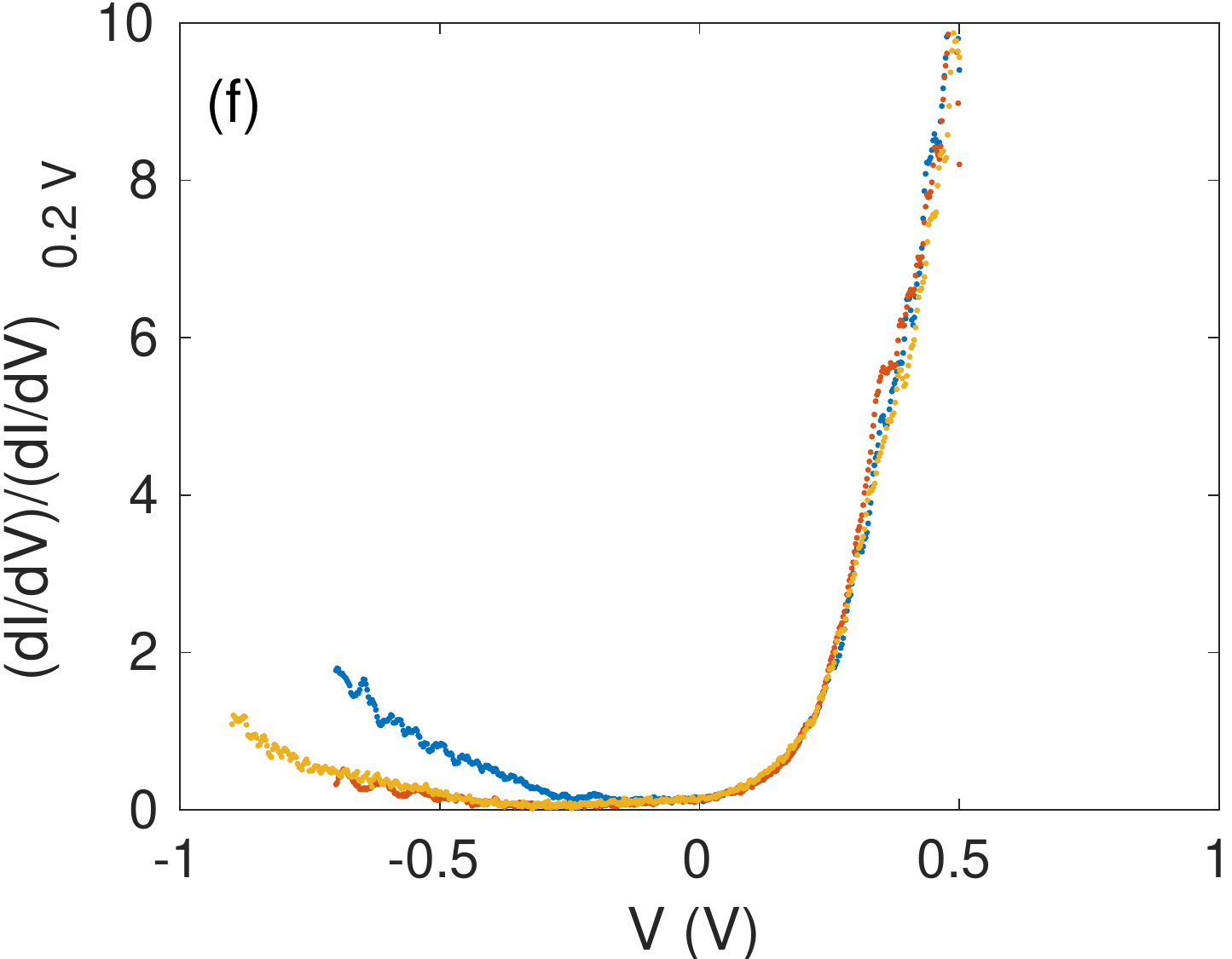}

\caption{\label{fig:a250}(a) $800\times 800$ nm$^2$ STM image of a film deposited at substrate temperature 280 \textdegree C at argon pressure 0.3 Torr and annealed in UHV conditions during 1 hour;
(b) STM image of the film annealed at 300 \textdegree C;
(c) STM image of a typical defect; (d) surface degradation region (color scales are in nanometers).
(e) LDOS of a flat surface.
(f) LDOS of reconstructed regions.}
\end{figure}

Fig. \ref{fig:a250}(a) shows STM image of $800\times 800$ nm$^2$ region of a film deposited at argon pressure 0.3 Torr. The film consists of islands whose thickness is a multiple of QL thickness (Fig. \ref{fig:a250}(b)). Some islands coalescent forming a percolative crystalline network. The island height correlates with their width, in agreement with the Gibbs-Wulff theorem. In the presence of the macroscopic substrate defects (folds, steps), the film grows along such defects (Fig.~\ref{fig:a250}(b)).
Atomic structure of terrace surfaces has hexagonal symmetry, with triangular defects  characteristic of  Bi$_2$Te$_3$ (Fig.~\ref{fig:a250}(c)). In addition, we observed small regions of surface degradation accompanied by superstructure development (Fig.~\ref{fig:a250}(d)) similar to the ones observed in films deposited at lower temperatures and annealed at 300 \textdegree C (see Fig.~\ref{fig:a50}(c), Fig.~\ref{fig:a150}(d)). The LDOS on such surfaces (Fig.~\ref{fig:a250}(f)) does not correspond to LDOS observed on undamaged surface 
 and resembles the LDOS of Bi-terminated cleaved surface (Fig.~\ref{fig:STS}(i)).

\section{Discussion}
As it can be seen from the results, thin polycrystalline Bi$_2$Te$_3$ films on the surface of pyrolytic graphite can be obtained by the PLD method in a surprisingly wide range of substrate and subsequent annealing temperatures.  Films deposited in vacuum from a target with an atomic ratio of components Bi/Te = 1/5 at a substrate temperature of %50–305
60-240~\textdegree C are polycrystalline with grain sizes of 10–50 nm and a random orientation of crystallites.  Such polycrystalline films may be of interest for thermoelectric applications.

An interesting behavior is observed for  films deposited in vacuum at a substrate temperature of 60 \textdegree C. The crystallites in these films are more ordered than for films deposited at higher temperatures. Annealing of these films at temperatures above 200 \textdegree C leads to the formation of a quasi-continuous wavy structure  resembling potato chips. The appearance of such a structure can be explained by the accretion of crystallites with different inclination relative to the substrate surface and opens up a possibility to study the effect of very strong crystal distortion on its electronic properties. This effect is not observed in films deposited at 175 and 240 \textdegree C apparently because of lack of tellurium in as-deposited films. %in accordance with the results of EDS analysis.   

A common feature of all the films is a noticeable sublimation of crystallites during annealing at temperatures of 300 \textdegree C and higher.
After complete disappearance of the film, a disordered layer remains on the substrate, which is not expected for the case of vdW epitaxy. We attribute the appearance of this layer to the presence of high-energy ions during laser sputtering of the target, which damage the substrate.

To eliminate the destructive effect of high-energy ions on the substrate, we carried out experiments on the deposition of films in an argon atmosphere at a pressure of 0.3 Torr. The small  value of the mean free path at this pressure (less than 1 mm) ensures the absence of high-energy ion collisions with the substrate. The films obtained as a result of such deposition and subsequent annealing at $T_a\gtrsim 200$ \textdegree C turn out to be highly crystalline, and after their sublimation, an intact atomic structure of HOPG is recovered.   Thus, the vdW epitaxy with PLD is only possible when laser ablation occurs in inert gas atmosphere.

PLD in argon atmosphere at moderate substrate temperatures also results in a more congruent material transfer from the target. In our case, this corresponds to a high Te content, with the excess Te spreading uniformly or forming large crystallites depending on the substrate temperature during deposition. Annealing at $T_a\gtrsim 200$ \textdegree C results in evaporation of the excess Te. 

The energy structure of the films corresponds to the structure of the topological insulator Bi$_2$Te$_3$.
In particular, STS spectra of the first three terraces of the film correspond to calculated LDOS of 1QL, 2QL and 3QL respectively. The heights of the second and third layers correspond with the expected height of 1 QL, while surprisingly the height of the first layer is larger and indicates the presence of a buffer layer.

We have also observed that annealing at  $T_a\approx 300$ \textdegree C results in development of a hexagonal atomic superstructure with a 4 nm period. The STS measurements indicate the presence of an energy gap at the Fermi energy, as it is expected in the case of the CDW transition. The origin of this CDW-like reconstruction requires further study.

\section{Conclusion}
In conclusion, the morphology and chemical composition of nanometer-scale thickness Bi-Te system  films grown by PLD method on HOPG substrates depend on deposition and  annealing conditions and can vary from a set of grains with different orientations to an epitaxial film. We  demonstrate that PLD in vacuum results in damage of the substrate and therefore cannot be used for vdW epitaxy. Instead, PLD at a moderate inert gas pressure followed by annealing at temperatures $T_a\approx 200$-250 \textdegree C enables vdW epitaxy of ultrathin crysalline films of Bi$_2$Te$_3$. The nature of a CDW-like surface reconstruction observed on the initial stage of film sublimation needs further study.

{\bf Acknowledgments.} The SEM measurements were performed using equipment of the Shared Facility Center at P. N. Lebedev Physical Institute of RAS. The computations in this study were performed using computational  resources at  the  Joint  Supercomputer Center, Russian Academy of Science. The financial support from the state task and RFBR (grant \#19-02-00593) is acknowledged. 

\bibliographystyle{unsrtnat}
%\bibliography{custom-2}

% Loading bibliography database
\bibliography{PLD_BixTey.bib}

\end{document}